\documentclass[11pt]{article}

\newlength{\actualtopmargin}
\newlength{\actualsidemargin}
\usepackage[tbtags]{amsmath}
\usepackage{HKamsthm}
\usepackage{amssymb}
\usepackage{amsfonts}
  \theoremstyle{plain}
  \newtheorem{theorem}{Theorem}
  \newtheorem{lemma}[theorem]{Lemma}
  \newtheorem{corollary}[theorem]{Corollary}

  \theoremstyle{definition}
  \newtheorem{definition}[theorem]{Definition}

  \theoremstyle{remark}
  
  \theoremstyle{plain}
  \newtheorem*{theorem*}{Theorem}
  \newtheorem*{lemma*}{Lemma}
  \newtheorem*{corollary*}{Corollary}
  \newtheorem*{proposition*}{Proposition}
  \newtheorem*{claim*}{Claim}
  \renewcommand{\qedsymbol}{$\square$}

\newenvironment{step}
  {
    \begin{enumerate}

  }
  {\end{enumerate}}

\newenvironment{algorithm*}[1]
  {
    \begin{center}
      \hrulefill\\
      \textbf{#1}
  }
  {
    \vspace{-1\baselineskip}
    \hrulefill
    \end{center}
  }

\newenvironment{protocol*}[1]
  {
    \begin{center}
      \hrulefill\\
      \textbf{#1}
  }
  {
    \vspace{-1\baselineskip}
    \hrulefill
    \end{center}
  }

\newcommand{\sfB}{\mathsf{B}}

\newcommand{\sfM}{\mathsf{M}}
\newcommand{\sfN}{\mathsf{N}}

\newcommand{\sfP}{\mathsf{P}}
\newcommand{\sfQ}{\mathsf{Q}}
\newcommand{\sfR}{\mathsf{R}}

\newcommand{\sfV}{\mathsf{V}}

\newcommand{\sfX}{\mathsf{X}}
\newcommand{\sfY}{\mathsf{Y}}

\newcommand{\classfont}{\mathrm}

\newcommand{\NP}{\classfont{NP}}

\newcommand{\PSPACE}{\classfont{PSPACE}}
\newcommand{\EXP}{\classfont{EXP}}
\newcommand{\NEXP}{\classfont{NEXP}}

\newcommand{\MIP}{\classfont{MIP}}
\newcommand{\QIP}{\classfont{QIP}}
\newcommand{\QMIP}{\classfont{QMIP}}
\newcommand{\publicQMIP}{\QMIP_{\mathrm{pub}}}

\newcommand{\QMAM}{\classfont{QMAM}}

\newcommand{\poly}{\mathrm{poly}}

\newcommand{\init}{\mathrm{init}}
\newcommand{\illegal}{\mathrm{illegal}}
\newcommand{\final}{\mathrm{final}}
\newcommand{\acc}{\mathrm{acc}}
\newcommand{\rej}{\mathrm{rej}}

\newcommand{\bra}[1]{\langle #1 \vert}

\newcommand{\ket}[1]{\vert #1 \rangle}

\newcommand{\ketbra}[1]{\vert #1 \rangle \langle #1 \vert}

\newcommand{\conjugate}[1]{#1^{\dagger}}

\newcommand{\norm}[1]{\Vert #1 \Vert}

\newcommand{\bignorm}[1]{\bigl\Vert #1 \bigr\Vert}

\newcommand{\abs}[1]{\vert #1 \vert}

\newcommand{\bigabs}[1]{\bigl\vert #1 \bigr\vert}

\newcommand{\ceil}[1]{\lceil #1 \rceil}

\newcommand{\function}[3]{{#1 \colon #2 \rightarrow #3}}

\newcommand{\Integers}{\mathbb{Z}}
\newcommand{\Nonnegative}{\Integers^{+}}

\newcommand{\Binary}{{\{ 0, 1 \}}}

\newcommand{\ignore}[1]{}

\begin{document}

\sloppy


\title{\Large
  \textbf{
    Using Entanglement in Quantum Multi-Prover Interactive Proofs
  }\\
}

\author{
Julia Kempe\thanks{Work partly done while at LRI, Univ. de Paris-Sud, Orsay. Partially supported by the
European Commission under the Integrated Project Qubit Applications (QAP) funded by the IST directorate as
Contract Number 015848, by an Alon Fellowship of the Israeli Higher Council of Academic Research and by a
grant of the Israeli Science Foundation.}\\School of Computer Science
\\Tel Aviv University\\ Tel Aviv, Israel\\
 \and
 Hirotada Kobayashi\thanks{Supported by
the Strategic Information and Communications R\&D Promotion Programme No. 031303020 of the Ministry of
Internal Affairs and Communications of Japan and the Grant-in-Aid for Scientific Research (B) No. 18300002 of
the Ministry of Education, Culture, Sports, Science and Technology of Japan.
}\\
 Principles of Informatics Research Division\\
  National Institute of Informatics\\
  Tokyo,
  Japan\\
 \and
 Keiji Matsumoto$^\dagger$\\
 Principles of Informatics Research Division\\
  National Institute of Informatics\\
  Tokyo,
  Japan\\
 \and
Thomas Vidick\thanks{Work partly done while at LRI, Univ. de Paris-Sud, Orsay and DI, \'Ecole Normale Sup\'erieure, Paris.}\\
Computer Science Division\\ University of California, Berkeley\\
USA
 }
\date{\today}

\maketitle
\thispagestyle{empty}
\pagestyle{plain}
\setcounter{page}{0}

\renewcommand{\thefootnote}{\fnsymbol{footnote}}

\renewcommand{\thefootnote}{\arabic{footnote}}


\begin{abstract}
The central question in quantum multi-prover interactive proof systems is whether or not entanglement shared
between provers affects the verification power of the proof system. We study for the first time
\emph{positive} aspects of prior entanglement and show  that  entanglement \emph{is} useful even for
\emph{honest} provers. We show how to use shared entanglement to parallelize any multi-prover quantum
interactive proof system to a {\em one-round} system with {\em perfect completeness}, with one extra prover.
Alternatively, we can also parallelize to a three-turn system with the same number of provers, where the
verifier only broadcasts the outcome of a coin flip. This ``public-coin" property is somewhat surprising,
since in the classical case public-coin multi-prover interactive proofs are equivalent to single prover ones.

\end{abstract}

\clearpage


\section{Introduction}
\label{Section: introduction}

Multi-prover interactive proof systems are a central notion in theoretical computer science. An important
generalization of interactive proof systems~\cite{GolMicRac89SIComp, Bab85STOC}, they were originally
introduced in~\cite{BenGolKilWig88STOC} in a cryptographic context. Later it was
shown~\cite{BabForLun91CC,ForRomSip94TCS} that the class $\MIP$ of languages having a multi-prover
interactive proof system is equal to $\NEXP$, which led to the development of the theory of inapproximability
and probabilistically checkable proofs~\cite{FeiGolLovSafSze96JACM, AroSaf98JACM, AroLunMotSudSze98JACM}.

In a multi-prover interactive proof system, a verifier communicates with several provers, which do not
communicate with each other. One of the central challenges in this area is to understand the power of {\em
quantum} multi-prover interactive proof systems (QMIP systems). In particular, the major open question is how
{\em entanglement} shared among the provers affects these systems. This question is unique to the quantum
world, since the related classical resource of shared randomness is known not to affect the power of such
systems. It is not even clear whether entanglement {\em increases} or {\em decreases} the verification power
of QMIP systems. On one hand, using entanglement, dishonest provers might cheat more easily, thereby breaking
the soundness of the system. On the other hand, the increased power that entanglement gives to honest provers
could be harnessed by the verifier, increasing the expressivity of the proof system.

To the best of our knowledge, all previous results in this area (see below) have focused on the former case,
studying the {\em negative} effects of entanglement, i.e., whether or not \emph{dishonest} entangled provers
can break proof systems that are sound for any dishonest \emph{unentangled} provers. Our work is the first to
focus on the {\em positive} aspects of entanglement, where shared entanglement may be advantageous to {\em
honest} provers.

\subsection{Previous and related work}

Kobayashi~and~Matsumoto~\cite{KobMat03JCSS} introduced QMIP systems with a quantum verifier, and proved that
the class of languages having a quantum multi-prover interactive proof system is equal to $\NEXP$ when the
provers do not share any prior entanglement, and is contained in $\NEXP$ when they share at most polynomially
many entangled qubits. Cleve,~H{\o}yer,~Toner,~and~Watrous~\cite{CleHoyTonWat04CCC} studied multi-prover
interactive proof systems in which the verifier remains classical but provers may initially share
entanglement, and presented several protocols for which shared EPR pairs can increase the power of dishonest
provers. They also proved that the class of languages having some restricted version of multi-prover
interactive proof system, denoted by $\oplus\MIP^*(2,1)$, is contained in $\EXP$ when provers are allowed to
share prior entanglement (Wehner~\cite{Weh06STACS} improved the upper bound to ${\QIP(2)}$, the class of
languages having a two-message quantum interactive proof system), which is in stark contrast to the
corresponding class ${\oplus\MIP(2,1)}$ without prior entanglement, which is equal to $\NEXP$\footnote{for
some two-sided bounded error}. Very recently,
Kempe,~Kobayashi,~Matsumoto,~Toner,~and~Vidick~\cite{KemKobMatTonVid07arXiv} gave limits on the cheating
power of dishonest entangled provers in some quantum and classical multi-prover interactive proof systems, by
showing how such proof systems can be ``immunized" against the use of entanglement by dishonest provers.
Ito,~Kobayashi,~Preda,~Sun,~and~Yao~\cite{yao:tsirelson} and
Cleve,~Gavinsky,~and~Jain~\cite{CleGavJai07arXiv} also gave limits on the cheating power of entangled provers
for some classical multi-prover interactive proof systems.

All these studies focus only on the \emph{negative} aspects of prior entanglement, i.e., whether or not
\emph{dishonest} but prior-entangled provers can break the soundness of the proof system.

\subsection{Our Results}

This paper studies the \emph{positive} aspects of prior entanglement and shows a number of general properties
of QMIP systems, extensively using prior entanglement for \emph{honest} provers. This gives the first
evidence that prior entanglement \emph{is} useful even for honest provers. Our main theorem states that any
quantum $k$-prover interactive proof system that may involve polynomially many rounds can be parallelized to
a \emph{one-round} quantum ${(k+1)}$-prover interactive proof system of {\em perfect} completeness and such
that the gap between completeness and soundness accepting probabilities is still bounded by an
inverse-polynomial.

To state our results more precisely, let ${\QMIP(k, m, c, s)}$ denote the class of languages having an
$m$-turn quantum $k$-prover interactive proof system with completeness at least $c$ and soundness at most
$s$, where provers are allowed to share an arbitrary amount of entanglement.
We call the difference $c-s$ the ``gap" in this paper. As commonly used in
classical multi-prover interactive proofs we use the term ``round'' to describe an interaction consisting of
questions from the verifier followed by answers from the provers. We use the term ``turn" for messages sent
in one direction. One round consists of two turns: a turn for the verifier and a turn for the provers. Let
$\poly$ and $\poly^{-1}$ be the sets of all polynomially bounded functions and all inverse-polynomial
functions, respectively. Throughout this paper we assume that the number $m$ of turns and the number $k$ of
provers are functions in $\poly$ with respect to the input size, and that completeness $c$ and soundness $s$
are functions of the input size $\function{c, s}{\Nonnegative}{[0,1]}$. Then we show the following main
theorem.

\begin{theorem}
For any $k,m \in \poly$ and $c,s$ satisfying ${c - s \in \poly^{-1}}$ there exists a function ${p \in \poly}$
such that
 ${
  \QMIP(k, m, c, s)
  \subseteq
  \QMIP \left( k+1, 2, 1, 1 - \frac{1}{p} \right)
}$. \label{Theorem: parallelization to 1 round}
\end{theorem}

Since it is easy to amplify the success probability without increasing the number of rounds by running
multiple instances of a proof system in parallel using a different set of provers for every instance, the
above theorem shows that one-round (i.e., two-turn) QMIP systems are as powerful as general QMIP systems.

\begin{corollary}
For any $k,m \in \poly$ and $c,s$ satisfying ${c - s \in \poly^{-1}}$, and $ p \in \poly$, there exists $ k'
\in \poly$ such that ${
  \QMIP(k, m, c, s)
  \subseteq
  \QMIP(k', 2, 1 , 2^{-p})
}$.\label{Corollary: 1 round QMIP = general QMIP}
\end{corollary}

The proof of our main theorem  comes in three parts, corresponding to Sections \ref{Section: QMIP with
Perfect Completeness Equals General QMIP}, \ref{Subsection: Parallelizing to Three Turns}, and
\ref{Section:part3}. The first part shows how to convert any QMIP system with two-sided bounded error into
one with one-sided bounded error of perfect completeness without changing the number of provers.
The second part shows that any QMIP system with polynomially many turns can be parallelized to one with only
three turns (messages from the provers followed by questions from the verifier followed by responses from the
provers) in which the gap between completeness and soundness is still bounded by an inverse-polynomial. Again
the number of provers remains the same in this transformation. Finally, the third part shows that any
three-turn QMIP system with sufficiently large gap can be converted into a two-turn (i.e., one-round) QMIP
system with inverse-polynomial gap, by adding an extra prover.

Similar statements to our first and second parts have already been shown by Kitaev and
Watrous~\cite{KitWat00STOC} for {\em single-prover} quantum interactive proofs. Their proofs, however,
heavily rely on the fact that a single quantum prover can apply arbitrary operators over all the space except
for the private space of the verifier. This is not the case any more for quantum multi-prover interactive
proofs, since now a quantum prover cannot access the qubits in the private spaces of the other quantum
provers, in addition to those in the private space of the verifier. Hence new methods are required for the
multi-prover case.

To transform proof systems so that they have perfect completeness, our basic idea is to use the quantum
rewinding technique developed for quantum zero-knowledge proofs by Watrous~\cite{Wat06STOC}, but in a
different way. In our case we use it to ``rewind'' an unsuccessful computation that would result in rejection
into a successful one. To apply the quantum rewinding technique, we first modify the proof system so that the
honest provers can convince the verifier with probability exactly $\frac{1}{2}$ using some initial shared
state and moreover no other initial shared state achieves a higher acceptance probability. This initial
shared state corresponds to the auxiliary input in the case of quantum zero-knowledge proofs, and thus, as in
that scenario, the sequence of forward, backward, and forward executions of the protocol achieves perfect
completeness. The obvious problem of this construction lies in proving soundness, as the dishonest provers
may not use the same strategies for all of the three executions of the proof system. To settle this, we
design a simple protocol that tests if the second backward execution is indeed a backward simulation of the
first forward execution. The verifier performs with equal probability either the original rewinding protocol
or this invertibility test without revealing which test the provers are undergoing. This forces the provers
to use essentially the same strategies for the first two executions of the protocol, which is sufficient to
bound the soundness. As a result we prove the following.

\begin{theorem}
For any $k,m \in \poly$ and $c,s$ satisfying ${c - s \in \poly^{-1}}$, and $ p \in \poly$, there exists $m'
\in \poly$ such that ${
  \QMIP(k, m, c, s)
  \subseteq
  \QMIP(k, m', 1, 2^{-p})
}$.\label{Theorem: QMIP with perfect completeness}
\end{theorem}
For the parallelization to three turns, our approach is to first show that any  QMIP system with sufficiently
large gap can be converted into another  QMIP system with the same number of provers, in which the number of
rounds (turns) becomes almost half of that in the original proof system. The proof idea is that the verifier
in the first turn receives the snapshot state from the original system after (almost) half of turns have been
executed, and then with equal probability executes either a forward-simulation or a backward-simulation of
the original system from that turn on. Honest provers only have to simulate the original system to convince
the verifier, while any strategy of dishonest provers with unallowable high success probability would lead to
a strategy of dishonest provers in the original system that contradicts the soundness condition. By
repeatedly applying this modification, together with Theorem \ref{Theorem: QMIP with perfect completeness} as
preprocessing, we can convert any  QMIP system into a three-turn  QMIP system with the same number of provers
that still has an inverse polynomial gap.

\begin{theorem}
For any $k,m \in \poly$ and $c,s$ satisfying ${c - s \in \poly^{-1}}$, there exists $p \in \poly$ such that
 ${
  \QMIP(k, m, c, s)
  \subseteq
  \QMIP \left( k, 3, 1, 1 - \frac{1}{p} \right)
}$. \label{Theorem: parallelization to 3 turns}
\end{theorem}

For ${k = 1}$, this gives an alternative proof of the parallelization theorem due to
Kitaev~and~Watrous~\cite{KitWat00STOC} for single-prover quantum interactive proofs. It is interesting to
note that our parallelization method does not need the controlled-swap test at all, while it \emph{is} the
key test in the Kitaev-Watrous parallelization method. Another point worth mentioning in our method is that,
at every time step of our parallelized protocol, the whole system has only one snapshot state of the original
system. This is in contrast to the fact that the whole system has to simultaneously treat many snapshot
states in the Kitaev-Watrous method. The merit of our method is, thus, that we do not need to treat the
possible entanglement among different snapshot states when analyzing soundness, which may be a main reason
why our method works well even for the multi-prover case. Moreover, our method is more space-efficient than
the Kitaev-Watrous method, in particular when we parallelize a system with polynomially many rounds.

To prove the third part, we will take a detour by proving that
\begin{itemize}
\item[(i)] any three-turn QMIP system with sufficiently large gap can be modified to a three-turn
\emph{public-coin} QMIP system with the same number of provers and a gap of roughly similar order of
magnitude,

\item[(ii)] any three-turn public-coin QMIP system can be converted into a two-turn QMIP system without
changing completeness and soundness, by adding one extra prover.
\end{itemize}

The notion of public-coin QMIP systems we use is a natural generalization of public-coin quantum interactive
proofs in the single-prover case introduced by Marriott~and~Watrous~\cite{MarWat05CC}. The corresponding
complexity class is denoted by $\publicQMIP(k,m,c,s)$ in this paper.  Intuitively, at every round, a
public-coin quantum verifier flips a fair classical coin at most polynomially many times, and then simply
broadcasts the result of these coin-flips to all the provers. Property (i) is a generalization of the result
by Marriott~and~Watrous~\cite{MarWat05CC} to the multi-prover case, whereas property (ii) is completely new.
The idea to prove (ii), assuming that the number of provers in the original proof system is $k$, is to send
questions only to the first $k$ provers in the new $(k+1)$-prover system, requesting the original second
messages from the $k$ provers in the original system. The verifier expects to receive from the ${(k+1)}$-st
prover the original first messages from the $k$ provers in the original system without asking any question to
that prover. The public-coin property of the original system implies the nonadaptiveness of the messages from
the verifier, which is essential to prove (ii). In fact, there is a way to directly prove the third part,
but our detour enables us to show another two important properties of QMIP systems. Specifically, property
(i) essentially proves the equivalence of \emph{public-coin} quantum $k$-prover interactive proofs and
general quantum $k$-prover interactive proofs, for any $k$.

\begin{theorem}
For any ${k, m \in \poly}$ and $c,s$ satisfying ${c - s \in \poly^{-1}}$, and ${p \in \poly}$, there exists
${m' \in \poly}$ such that ${
  \QMIP(k, m, c, s)
  \subseteq
  \publicQMIP(k, m', 1, 2^{-p})
}$. \label{Theorem: public-coin QMIP = general QMIP}
\end{theorem}

Note that in the classical case, public-coin multi-prover interactive proofs are only as powerful as
single-prover interactive proofs: because every prover receives the same question from the verifier it means
that every prover knows how other provers will behave and the joint strategy of the provers can therefore be
simulated by a single prover. Hence, these systems cannot be as powerful as general classical multi-prover
interactive proofs unless ${\NEXP = \PSPACE}$. In contrast, our result shows that in the quantum case,
public-coin QMIP systems {\em are} as powerful as general QMIP systems. The non-triviality of public-coin
QMIP systems may be explained as follows: even if every quantum prover knows how other quantum provers will
behave, still each quantum prover can apply only local transformations over a part of some state that may be
entangled among the provers, which is not enough to simulate every possible strategy a single quantum prover
could follow.

Property (ii) for the case ${k=1}$ implies that any language in $\QIP$ (and thus in $\PSPACE$) has a
\emph{two-prover one-round} quantum interactive proof system of perfect completeness with exponentially small
error in soundness, since any language in $\QIP$ has a three-message public-coin quantum interactive proof
system of perfect completeness with exponentially small error in soundness~\cite{MarWat05CC}.

\begin{corollary}
For any $p \in \poly$, ${\QIP \subseteq \QMIP(2, 2, 1, 2^{-p}) }$ (and thus ${\PSPACE \subseteq \QMIP(2, 2,
1, 2^{-p}) }$).\label{Corollary: QIP is in QMIP(2,2,1,s)}
\end{corollary}

In the classical case a similar statement to the last corollary was shown by
Cai,~Condon,~and~Lipton~\cite{CaiConLip94JCSS} (and the stronger statement that two-prover one-round
interactive proofs are as powerful as general multi-prover interactive proofs was shown later by
Feige~and~Lov\'asz~\cite{FeiLov92STOC}). All these results are, however, not known to hold under the
existence of prior entanglement among the provers. Before our result, it has even been open if $\PSPACE$ has
a two-prover one-round quantum interactive proof system. (Very recently, Kempe et
al.~\cite{KemKobMatTonVid07arXiv} succeeded in proving that the classical two-prover one-round interactive
proof system for $\PSPACE$ in Ref.~\cite{CaiConLip94JCSS} is sound in a weak sense against any pair of
dishonest prior-entangled provers: soundness is bounded away from one by an inverse-polynomial. Their result
is incomparable to ours since on one hand we have a much stronger soundness condition, and on the other both
the verifier and the  honest provers must be quantum. In contrast, in Ref.~\cite{KemKobMatTonVid07arXiv} both
of them just follow a classical protocol.)

Finally, we stress again that our constructions extensively use the prior shared entanglement of the provers
in a positive sense. In particular, even if the honest provers in the original proof system do not need any
prior entanglement at all, the honest provers in the constructed proof system do need prior entanglement in
many cases. Most of the properties proved in this paper (Theorems~\ref{Theorem: parallelization to 1 round}
and \ref{Theorem: public-coin QMIP = general QMIP} and Corollary \ref{Corollary: QIP is in QMIP(2,2,1,s)} in
particular) are not known to hold when considering only initially unentangled honest provers, and thus give
first evidence that sharing prior entanglement may be advantageous even to honest provers.




\section{Preliminaries}
\label{Section: Preliminaries}

We assume that the reader is familiar with the quantum formalism, including the quantum circuit model and
definitions of mixed quantum states (density operators) and fidelity (all of which are discussed in detail in
Refs.~\cite{NieChu00Book, KitSheVya02Book}, for instance). This section summarizes some of the notions and
notations that are used in this paper, reviews the model of quantum multi-prover interactive proof systems
and introduces the notion of \emph{public-coin} quantum multi-prover interactive proof systems.




As in earlier work~\cite{Wat03TCS,KitWat00STOC,KobMat03JCSS}, we define QMIP systems in terms of quantum
circuits. It is assumed that our circuits consist of unitary gates, which is sufficient since non-unitary and
unitary quantum circuits are equivalent in computational power~\cite{AhaKitNis98STOC}. To avoid unnecessary
complication, however,  in the subsequent sections the descriptions of protocols often include non-unitary
operations (measurements). Even in such cases, it is always possible to construct unitary quantum circuits
that essentially achieve the same outcome. A notable exception is in the definition of the public-coin
quantum verifier, where we want to define the public coin-flip to be a classical operation. This requires a
non-unitary operation for the verifier, the (classical) public coin-flip.

When proving statements that involve the perfect-completeness property, we assume that our universal gate set
satisfies some conditions, which may not hold with an arbitrary universal gate set. Specifically, we assume
that the Hadamard transformation and any classical reversible transformations are exactly implementable in
our gate set. Note that this condition is satisfied by most of the standard gate sets including the Shor
basis~\cite{Sho96FOCS} consisting of the Hadamard gate, the controlled-$i$-phase-shift gate, and the Toffoli
gate, and thus, we believe that this condition is not restrictive. We stress that most of our main statements
do hold with an arbitrary choice of universal gate set (the completeness and soundness conditions may become
worse by negligible amounts in some of the claims, which does not affect the final main statements).

All Hilbert spaces in this paper are of dimension a power of two, spanned by qubits. We will use the
following property of fidelity.

\begin{lemma}[\cite{SpeRud02PRA,NaySho03PRA}]
For any density operators $\rho, \sigma, \xi $ over a Hilbert space $\cal H$, ${F(\rho, \sigma)^2 + F(\sigma,
\xi)^2 \leq 1 + F(\rho, \xi)}$.
\label{Lemma: F(a,b)^2 + F(b,c)^2 < 1 + F(a,c)}
\end{lemma}


\paragraph{Quantum Multi-Prover Interactive Proof Systems (QMIP systems):}

Throughout this paper $k$ and $k'$  denote the number of provers and $m,m'$ denote the number of turns. All
of these are from the set of polynomially bounded functions in the input size $|x|$, denoted by $\poly$.
Further, $c$ and $s$ denote functions of the input size into $[0,1]$ corresponding to completeness and
soundness. For notational convenience in what follows  we will omit the arguments of these functions.

A quantum $k$-prover interactive proof system consists of a verifier $V$ with private quantum register $\sfV$
and $k$ {\em provers} $P_1,\ldots ,P_k$ with private quantum registers $\sfP_1,\ldots ,\sfP_k$, as well as
quantum message registers ${\sfM}_1, \ldots ,{\sfM}_k$, which without loss of generality are assumed to have
the same number of qubits, denoted by $q_{\sfM}$. One of the private qubits of the verifier is designated as
the {\em output qubit}. At the beginning of the protocol, all the qubits in ${(\sfV, \sfM_1, \ldots,
\sfM_k)}$ are initialized to $\ket{0\cdots 0}$, and the qubits in ${(\sfP_1, \ldots, \sfP_k)}$ are in some
\emph{a priori shared state} $\ket{\Phi}$ prepared by the provers in advance (and hence possibly entangled),
which w.l.o.g. can be assumed to be pure. No direct communication between the provers is allowed after that.
The protocol consists of alternating turns of the provers and of the verifier, starting with the verifier, if
$m$ is even, and with the provers otherwise. At a turn of the verifier, $V$ applies some polynomial-time
circuit to the qubits in ${(\sfV, \sfM_1, \ldots, \sfM_k)}$, and then sends each register $\sfM_i$ to prover
$P_i$. At a turn of the provers each prover $P_i$ applies some transformation to the registers
$(\sfP_i,\sfM_i)$ for $1 \leq i \leq k$ and sends $\sfM_i$ back to the verifier. The last turn is always a
turn for the provers. After the last turn the verifier applies a polynomial-time circuit to the qubits in
${(\sfV, \sfM_1, \ldots, \sfM_k)}$, and then measures the output qubit in the standard basis, accepting if
the outcome is $\ket{1}$ and rejecting otherwise.

Formally, an {\em $m$-turn polynomial-time quantum verifier $V$} for $k$-prover QMIP systems is a
polynomial-time computable mapping from input strings $x$ to a set of polynomial-time uniformly generated
circuits $\{ V^1,\ldots ,V^{\lceil m+1/2 \rceil} \}$, and a partition of the space on which they act into
registers $({\sfV},\sfM_1,\ldots ,{\sfM}_k)$, which consist of polynomially many qubits. Similarly an {\em
$m$-turn quantum prover ${P}$} is a mapping from $x$ to a set of circuits $\{ P^1,\ldots ,P^{\lceil m+1/2
\rceil} \}$ each acting on registers $({\sfP}, {\sfM})$. No restrictions are placed on the complexity of this
mapping or the size of $\sfP$. We will denote the $i$-th prover, his registers and transformations with a
subscript $i$. We will always assume that each prover $P_i$ is {\em compatible} with the verifier, i.e. that
the corresponding register ${\sfM}_i$ is the same for the verifier and the prover for $1 \leq i \leq k$.

The {\em protocol} $(V,P_1,\ldots ,P_k,\ket{\Phi})$ is the alternating application of prover's and verifier's
circuits to the state $\ket{0 \cdots 0}\otimes\ket{\Phi}$ in registers $(\sfV,\sfM_1,\ldots ,
\sfM_k,\sfP_1,\ldots ,\sfP_k)$. For odd $m$, circuits $P^1_1 \otimes \cdots \otimes P_k^1$, ${V^1}$, $P_1^2
\otimes \cdots \otimes P_k^2$, $V^2$ and so on are applied in sequence terminating with $V^{m+1/2}$. If $m$
is even, the sequence begins with $V^1$ followed by ${P}^1_1 \otimes \cdots \otimes {P}_k^1$ and so on up to
$V^{m+2/2}$. We say that $(V, P_1, \ldots, P_k, \ket{\Phi})$ accepts $x$ if the designated output qubit in
${\sfV}$ is measured in $\ket{1}$ at the end of the protocol and call the probability with which this happens
$p_{\acc}(x,V,P_1,\ldots,P_k,\ket{\Phi})$.

\begin{definition}
A language $L$ is in ${\QMIP(k, m, c, s)}$ iff there exists an $m$-turn polynomial-time quantum verifier $V$
for quantum $k$-prover interactive proof systems such that, for every input $x$:
\begin{description}
\item[\textnormal{(Completeness)}] if ${x \in L}$, there exist $m$-turn quantum provers ${P_1, \ldots,
P_{k}}$ and an a priori shared state $\ket{\Phi}$ such that $p_{\acc}(x,V,P_1,\ldots,P_k,\ket{\Phi})\geq{c}$,
 \item[\textnormal{(Soundness)}] if ${x \not\in L}$, for any $m$-turn quantum
provers ${P'_1, \ldots, P'_{k}}$ and any a priori shared state $\ket{\Phi'}$,
$p_{\acc}(x,V,P'_1,\ldots,P'_k,\ket{\Phi'})\leq {s}$.
\end{description}
\label{Definition: QMIP(k,m,c,s)}
\end{definition}
Next, we introduce the notions of \emph{public-coin} quantum verifier and \emph{public-coin} QMIP systems.
These are natural generalizations of the corresponding notions in the single-prover case introduced by
Marriott~and~Watrous~\cite{MarWat05CC}.
Intuitively, a quantum verifier for quantum multi-prover interactive proof systems is public-coin if, at each
of his turns, after receiving the message registers from the provers, he first flips a fair classical coin at
most a polynomial number of times, and then simply broadcasts the result of these coin-flips to all the
provers. No other messages are sent from the verifier to the provers. At the end of the protocol, the
verifier applies some quantum operation to the messages received so far, and decides acceptance or rejection.

Formally, an $m$-turn polynomial-time quantum verifier for $k$-prover interactive proof systems is {\em
public-coin} if each of the circuits $V^1, V^2, \ldots ,V^{\lceil m-1/2 \rceil}$ implements the following
procedure: $V$ receives the message registers $\sfM_i$ from the provers, stores them in his private space,
and then flips a classical fair coin at most $q_{\sfM}$ times to generate a public string $r_j$, records
$r_j$ in his private space, and broadcasts $r_j$ to all the provers. The circuit ${V^{\ceil{(m+1)/2}}}$ is
some unitary transformation controlled by all the recorded random strings $r_j$ for ${1 \leq j \leq
\ceil{(m-1)/2}}$. A QMIP system is public-coin if the associated verifier is public-coin, and we define
${\publicQMIP(k,m,c,s)}$ to be the class of languages in $\QMIP(k,m,c,s)$ with a public-coin verifier.



\section{QMIP with Perfect Completeness Equals General QMIP}
\label{Section: QMIP with Perfect Completeness Equals General QMIP}

In this section we prove Theorem \ref{Theorem: QMIP with perfect completeness}, showing that any QMIP system
with two-sided bounded error can be transformed into a one with one-sided bounded error of perfect
completeness without changing the number of provers. For the case of a single prover, this was shown by
Kitaev and Watrous \cite{KitWat00STOC}, but their proof relies on the single prover performing a global
unitary on the whole system, and therefore does not carry over to the multi-prover case (no prover has access
to the other prover's private spaces and the private space of each prover might be arbitrarily large, so we
cannot use the verifier to transfer those spaces from one prover to the other).

First, we introduce the notion of \emph{perfectly rewindable} QMIP systems.

\begin{definition}
Let ${s < \frac{1}{2}}$. A language $L$ has a perfectly rewindable $m$-turn quantum $k$-prover interactive
proof system with soundness at most $s$ iff there exists an $m$-turn polynomial-time quantum verifier $V$,
such that, for every input $x$:
\begin{description}
\item[\textnormal{(Perfect Rewindability)}] if ${x \in L}$, there exists a set of $m$-turn quantum provers
${P_1, \ldots, P_{k}}$ such that $\max_{\ket{\Phi}}p_{\acc}(x,V,P_1,\ldots,P_k,\ket{\Phi})=\frac{1}{2}$,
where the maximum is taken over all a priori shared states $\ket{\Phi}$ prepared by $P_1,\ldots ,P_k$.

\item[\textnormal{(Soundness)}] if ${x \not\in L}$, for any set of $m$-turn quantum provers ${P'_1, \ldots,
P'_{k}}$ and any a priori shared state $\ket{\Phi'}$,
$p_{\acc}(x,V,P'_1,\ldots,P'_k,\ket{\Phi'})\leq{s}$.\footnote{Note that both for completeness and soundness
we first fix the provers' transformations and then maximize over all a priori shared states, which hence have
a fixed dimension.}
\end{description}
\label{Definition: perfectly rewindable systems}
\end{definition}
We first show how to modify any general QMIP system (with some appropriate conditions on completeness and
soundness) to a perfectly rewindable one with the same $k$ and $m$.
\begin{lemma}
Let  ${c \geq \frac{1}{2} > s}$. Then, any language $L$ in ${\QMIP(k, m, c, s)}$ has a perfectly rewindable
$m$-turn quantum $k$-prover interactive proof system with soundness at most $s$. \label{Lemma: making QMIP
perfectly rewindable}
\end{lemma}

\begin{proof}
Let $L$ be a language in ${\QMIP(k, m, c, s)}$ and $V$ be the corresponding $m$-turn quantum verifier. We
slightly modify $V$ to construct another $m$-turn quantum verifier $W$ for a perfectly rewindable proof
system for $L$. The new verifier $W$, in addition to the registers of $V$, prepares another single-qubit
register ${\sfB}$, initialized to $\ket{0}$. For the first $m-2$ turns, $W$ simply simulates $V$. In the
$(m-1)$-st turn, a turn for the verifier, $W$ proceeds like $V$ would, but sends ${\sfB}$ to the first prover
in addition to the qubits $V$ would send in the original proof system. In the $m$-th turn the first prover is
requested to send ${\sfB}$ back to $W$, in addition to the qubits sent to $V$ in the original proof system.
Then $W$ proceeds for the final decision procedure like $V$ would, but accepts iff $V$ would have accepted
{\em and} ${\sfB}$ is in the state $\ket{1}$.\footnote{This protocol can be brought into the standard form
where only one qubit is measured to decide acceptance. Call ${\sfY}$ the register containing the designated
output qubit for $V$. $W$ adds a new single-qubit output register ${\sfX}$, initialized to $\ket{0}$. At the
end of the protocol $W$ performs a Toffoli gate on the qubits in ${\sfB, \sfY, \sfX}$ controlled by the
qubits in $({\sfB},{\sfY})$. Clearly ${\sfX}$ will contain $\ket{1}$ iff both ${\sfB}$ and ${\sfY}$ contain
$\ket{1}$.} Notice that $W$ accepts only if $V$ would have accepted, so the soundness is obviously at most
$s$ in the constructed proof system.

For perfect rewindability we slightly modify the protocol for honest provers in the case $x \in L$. Let
$\ket{\Phi^*}$ be the a priori shared state in the original proof system that maximizes the acceptance
probability for the original honest provers and let $p_{\max}$ be that maximal acceptance probability. The
new provers use $\ket{\Phi^*}$ as the a priori shared state and simulate the original provers except for the
last turn. The only difference is that in the last turn the first prover proceeds as $P_1$ would, {\em and}
applies a one-qubit unitary $T$ to the qubit in ${\sfB}$,
 \[ T:\ket{0} \rightarrow \sqrt{1 - \frac{1}{2 p_{\max}}} \ket{0} + \sqrt{\frac{1}{2 p_{\max}}} \ket{1}.\]
From the construction it is obvious that the maximum accepting probability is exactly equal to $\frac{1}{2}$
and that this maximum is achieved when the provers use the a priori shared state $\ket{\Phi^{\ast}}$.
\end{proof}

Now, we are ready to show the following lemma.

\begin{lemma}
Let ${c \geq \frac{1}{2}}$ and ${s < \frac{1}{25}}$. Then, ${
  \QMIP(k, m, c, s)
  \subseteq
  \QMIP \left( k, 3m, 1, \frac{1}{2} + 2 \sqrt{s} + \frac{5s}{2} \right)
}$. \label{Lemma: making QMIP perfect complete}
\end{lemma}

\begin{proof}
The intuitive idea behind the proof of this lemma, using Watrous' ``quantum rewinding technique", has already
been explained in the introduction. We add some more intuition before proceeding to the technical proof.
Using Lemma \ref{Lemma: making QMIP perfectly rewindable} we can assume that in the case of honest provers
($x \in L$) the acceptance probability with shared state $\ket{\Phi^*}$ is exactly $\frac{1}{2}$ and
furthermore that no other a priori shared state achieves higher acceptance probability. The acceptance
probability when the provers use any a priori shared state $\ket{\Phi}$ can be written as
$p_{\acc}=\|\Pi_{\acc} Q \ket{\Psi}\|^2=\|\Pi_{\acc} Q \Pi_{\init} \ket{\Psi}\|^2$, where
$\ket{\Psi}=\ket{0\cdots 0}_{(\sfV,\sfM_1,\ldots,\sfM_k)}\otimes \ket{\Phi}$, $Q$ is the unitary
transformation induced by the QMIP system just before the verifier's final measurement, ${\Pi_\init}$ is the
projection on $\ket{0\cdots 0}_{(\sfV,\sfM_1,\ldots,\sfM_k)}$ and $\Pi_{\acc}$ is the projection on $\ket{1}$
of the designated output qubit. In other words the state $\ket{\Psi^*}=\ket{0\cdots
0}_{(\sfV,\sfM_1,\ldots,\sfM_k)}\otimes \ket{\Phi^*}$ maximizes the expression
$$\max_{\ket{\Psi}} \langle { \Psi} |\Pi_{\init} Q^\dagger \Pi_{\acc} Q \Pi_{\init} |{ \Psi
}\rangle,$$ meaning that the matrix ${M = \Pi_{\init} \conjugate{Q} \Pi_{\acc} Q \Pi_{\init}}$ has maximum
eigenvalue $\frac{1}{2}$ with corresponding eigenvector $\ket{\Psi^*}$. Now we apply the quantum rewinding
technique by performing forward, backward, and forward executions of the proof system in sequence. Perfect
completeness follows from the fact that the initial state is an eigenvector of $M$ with the corresponding
eigenvalue exactly $\frac{1}{2}$, exactly as in the zero-knowledge scenario of \cite{Wat06STOC}.


The challenge of this construction lies in the proof of soundness. If the input is a no-instance, the maximum
eigenvalue of any matrix $M$ corresponding to our proof system is small. This shows that if the dishonest
provers are actually ``not so dishonest'', i.e., if they use the same strategies for all of the three
(forward, backward, and forward) executions of the original proof system, the acceptance probability is still
small. However, the problem arises when the dishonest provers change their strategies for some of the three
executions. To settle this, we design a simple protocol that tests if the backward execution is indeed a
backward simulation of the first forward execution. The verifier performs the original rewinding protocol or
this invertibility test uniformly at random without revealing which test the provers are undergoing. Honest
provers always pass this invertibility test, and thus perfect completeness is preserved. When the input is a
no-instance, this forces the provers to use approximately the same strategies for the first two executions of
the proof system, which is sufficient to bound the soundness.

We now proceed with the technical details. Let $L$ be a language in ${\QMIP(k, m, c, s)}$ and let $V$ be the
verifier in the perfectly rewindable $m$-turn quantum $k$-prover interactive proof system for $L$ as per
Lemma~\ref{Lemma: making QMIP perfectly rewindable}. We construct a $3m$-turn quantum verifier $W$ of a new
quantum $k$-prover interactive proof system for $L$. $W$ has the same registers as $V$ in the original proof
system, and performs one of two tests, which we call ``\textsc{Rewinding Test}'' and ``\textsc{Invertibility
Test}''. The exact protocol is described in Figure~\ref{Figure: Verifier's Protocol for Achieving Perfect
Completeness}, where for simplicity it is assumed that $m$ is even (the case in which $m$ is odd can be
proved in a similar manner).


\begin{figure}[h!]
\begin{algorithm*}{Verifier's Protocol for Achieving Perfect Completeness}
\begin{step}
\item Simulate the original verifier for the first $m$ turns.
 \item
  Choose ${b \in \Binary}$ uniformly at random.
  If ${b=0}$, move to the \textsc{Rewinding Test}
  described in Step~\ref{Rewinding Test},
  while if ${b=1}$, move to the \textsc{Invertibility Test}
  described in Step~\ref{Invertibility Test}.
\item
  (\textsc{Rewinding Test})\\
  \begin{step}
  \item
    Apply $V^{\frac{m}{2}+1}$ to the qubits in ${({\sfV}, {\sfM}_1, \ldots, {\sfM}_k)}$.
    Accept if the content of ${({\sfV}, {\sfM}_1, \ldots, {\sfM}_k)}$
    corresponds to an accepting state in the original proof system.
    Otherwise apply $\conjugate{(V^{\frac{m}{2}+1})}$ to the qubits in ${({\sfV}, {\sfM}_1, \ldots, {\sfM}_k)}$,
    and send ${\sfM}_i$ to the $i$th prover,
    for ${1 \leq i \leq k}$.
  \item
    For ${j = \frac{m}{2}}$ down to $2$, do the following:\\
    Receive ${\sfM}_i$ from the $i$th prover,
    for ${1 \leq i \leq k}$.
    Apply $\conjugate{(V^j)}$ to the qubits in ${({\sfV}, {\sfM}_1, \ldots, {\sfM}_k)}$,
    and send ${\sfM}_i$ to the $i$th prover,
    for ${1 \leq i \leq k}$.
  \item
    Receive ${\sfM}_i$ from the $i$th prover,
    for ${1 \leq i \leq k}$.
    Apply $\conjugate{(V^1)}$ to the qubits in ${({\sfV}, {\sfM}_1, \ldots, {\sfM}_k)}$.
    Perform a controlled-phase-flip:
    multiply the phase by $-1$ if all the
    qubits in ${({\sfV}, {\sfM}_1, \ldots, {\sfM}_k)}$ are in state $\ket{0}$.
    Apply $V_1$ to the qubits in ${({\sfV}, {\sfM}_1, \ldots, {\sfM}_k)}$,
    and send ${\sfM}_i$ to the $i$th prover,
    for ${1 \leq i \leq k}$.
    \label{controlled-phase-flip}
  \item
    For ${j = 2}$ to $\frac{m}{2}$, do the following:\\
    Receive ${\sfM}_i$ from the $i$th prover,
    for ${1 \leq i \leq k}$.
    Apply $V^j$ to the qubits in ${({\sfV}, {\sfM}_1, \ldots, {\sfM}_k)}$,
    and send ${\sfM}_i$ to the $i$th prover,
    for ${1 \leq i \leq k}$.
   \item
    Receive ${\sfM}_i$ from the $i$th prover,
    for ${1 \leq i \leq k}$.
    Apply $V^{\frac{m}{2}+1}$ to the qubits in ${({\sfV}, {\sfM}_1, \ldots, {\sfM}_k)}$.
    Accept if the content of ${({\sfV}, {\sfM}_1, \ldots, {\sfM}_k)}$
    corresponds to an accepting state in the original proof system,
    and reject otherwise.
  \end{step}
  \label{Rewinding Test}
\item
  (\textsc{Invertibility Test})\\
  \begin{step}
  \item
    Send ${\sfM}_i$ to the $i$th prover,
    for ${1 \leq i \leq k}$.
  \item
    For ${j = \frac{m}{2}}$ down to $2$, do the following:\\
    Receive ${\sfM}_i$ from the $i$th prover,
    for ${1 \leq i \leq k}$.
    Apply $\conjugate{(V^j)}$ to the qubits in ${({\sfV}, {\sfM}_1, \ldots, {\sfM}_k)}$,
    and send ${\sfM}_i$ to the $i$th prover,
    for ${1 \leq i \leq k}$.
  \item
    Receive ${\sfM}_i$ from the $i$th prover,
    for ${1 \leq i \leq k}$.
    Apply $\conjugate{(V^1)}$ to the qubits in ${({\sfV}, {\sfM}_1, \ldots, {\sfM}_k)}$.
    Accept if all the qubits in ${({\sfV}, {\sfM}_1, \ldots, {\sfM}_k)}$
    are in state $\ket{0}$,
    and reject otherwise.
  \end{step}
  \label{Invertibility Test}
\end{step}
\end{algorithm*}
\caption{Verifier's protocol for achieving perfect completeness} \label{Figure: Verifier's Protocol for
Achieving Perfect Completeness}
\end{figure}

{\em Completeness:} Assume the input $x$ is in $L$. From the original provers $P_1,\ldots ,P_k$ we design
honest provers $R_1,\ldots ,R_k$ for the constructed $3m$-turn system. Each new prover $R_i$ has the same
quantum register ${\sfP}_i$ as $P_i$ has, and the new provers initially share $\ket{\Phi^*}$. For the first
$m$ turns each $R_i$ simulates $P_i$. At the $(m+2j)$-th turn for $1 \leq j \leq \frac{m}{2}$, $R_i$ applies
$(P_i^{\frac{m}{2}-j+1})^\dagger$ (i.e. the inverse of the $(m-2j+2)$-nd turn of the original $P_i$) .
Finally, for the $(2m+2j)$-th turn for $1 \leq j \leq \frac{m}{2}$, $R_i$ again applies $P_i^{j}$.

It is obvious from this construction that the provers ${R_1, \ldots, R_k}$ can convince $W$ with certainty
when $W$ performs the \textsc{Invertibility Test}. We show that ${R_1, \ldots, R_k}$ can convince $W$ with
certainty even when $W$ performs the \textsc{Rewinding Test}. In short, this holds for essentially the same
reason that the quantum rewinding technique works well in the case of quantum zero-knowledge proofs, and we
will closely follow that proof.

For notational convenience, let ${\widetilde{P}^j = P_1^j \otimes \cdots \otimes P_k^j}$ for ${1 \leq j \leq
\frac{m}{2}}$, and let ${
  Q
  =
  V^{\frac{m}{2}+1}
  \widetilde{P}^{\frac{m}{2}}
  V^{\frac{m}{2}}
  \cdots
  \widetilde{P}^1
  V^1
}$. Recall that $M\ket{\Psi^*}=\frac{1}{2}\ket{\Psi^*}$ where $M=\Pi_{\init} \conjugate{Q} \Pi_{\acc} Q
\Pi_{\init}$. Define the unnormalized states $\ket{\phi_0}$, $\ket{\phi_1}$, $\ket{\psi_0}$, and
$\ket{\psi_1}$ by
\begin{align*}
  \ket{\phi_0}
  &=
  \Pi_{\acc} Q \ket{\Psi^*},
  &
  \ket{\phi_1}
  &=
  \Pi_{\rej} Q \ket{\Psi^*},
  &
  \ket{\psi_0}
  &=
  \Pi_{\init} \conjugate{Q} \ket{\phi_0},
  &
  \ket{\psi_1}
  &=
  \Pi_{\illegal} \conjugate{Q} \ket{\phi_0},
\end{align*}
where ${
  \Pi_{\illegal}
  =
  I_{(\sfV,\sfM_1,\ldots ,\sfM_k)}
  -
  \Pi_{\init}
}$ is the projection onto states orthogonal to $\ket{0\cdots 0}_{(\sfV,\sfM_1,\ldots ,\sfM_k)}$ and
$\Pi_{\rej}=I_{(\sfV,\sfM_1,\ldots ,\sfM_k)}-\Pi_{\acc}$. Then, noticing that ${\ket{\Psi^*} = \Pi_{\init}
\ket{\Psi^*}}$, we have
\[
\ket{\psi_0} = \Pi_{\init} \conjugate{Q} \Pi_{\acc} Q \ket{\Psi^*} = \Pi_{\init} \conjugate{Q} \Pi_{\acc} Q
\Pi_{\init} \ket{\Psi^*} = M \ket{\Psi^*} = \frac{1}{2} \ket{\Psi^*},
\]
and thus,
\[
\conjugate{Q} \ket{\phi_1} = \conjugate{Q} \Pi_{\rej} Q \ket{\Psi^*} = \ket{\Psi^*}-\conjugate{Q} \Pi_{\acc}
Q \ket{\Psi^*} = \ket{\Psi^*} - \conjugate{Q} \ket{\phi_0} = 2 \ket{\psi_0} - (\ket{\psi_0} + \ket{\psi_1}) =
\ket{\psi_0} - \ket{\psi_1}.
\]
Hence, the state just before the controlled-phase-flip in Step~3.3 when entering the \textsc{Rewinding Test}
is exactly
\[
\frac{1}{\norm{\ket{\phi_1}}} \conjugate{Q} \ket{\phi_1} = \frac{1}{\norm{\ket{\phi_1}}} (\ket{\psi_0} -
\ket{\psi_1}).
\]
Since ${\Pi_{\init} \ket{\psi_0} = \ket{\psi_0}}$ and ${\Pi_{\init} \ket{\psi_1} = 0}$, the
controlled-phase-flip changes the state to
\[
-\frac{1}{\norm{\ket{\phi_1}}} (\ket{\psi_0} + \ket{\psi_1}) = -\frac{1}{\norm{\ket{\phi_1}}} \conjugate{Q}
\ket{\phi_0}.
\]
Therefore, the state just after $V^{\frac{m}{2}+1}$ is applied in Step~3.5 is exactly
\[
-\frac{1}{\norm{\ket{\phi_1}}} Q \conjugate{Q} \ket{\phi_0} = -\frac{1}{\norm{\ket{\phi_1}}} \ket{\phi_0},
\]
and thus, the fact that ${\Pi_{\acc} \ket{\phi_0} = \ket{\phi_0}}$ implies that the verifier $W$ always
accepts in Step~3.5.

{\em Soundness:} Now suppose that the input $x$ is not in $L$. Let $R_1', \ldots ,R'_k$ be any $k$ provers
for the constructed $3m$-turn proof system, and let $\ket{\psi}$ be any a priori shared state. Let $R_i^{j}$
be the transformation that $R'_i$ applies at his $2j$-th turn, for $1 \leq i \leq k$ and $1\leq j \leq
\frac{3m}{2}$ and let $Z$ denote the controlled-phase-flip operator in Step~3.3. Call ${\widetilde{R}^j =
R_1^{j} \otimes \cdots \otimes R_k^{j}}$ for ${1 \leq t \leq \frac{3m}{2}}$, and define
\begin{align*}
  U_1
  &
  =
  \widetilde{R}^{\frac{m}{2}}
  V^{\frac{m}{2}}
  \cdots
  \widetilde{R}^2
  V^2
  \widetilde{R}^1
  V^1,
\\
  U_2
  &
  =
  \conjugate{(V^1)}
  \widetilde{R}^{m}
  \cdots
  \conjugate{(V^{\frac{m}{2}-1})}
  \widetilde{R}^{\frac{m}{2}+2}
  \conjugate{(V^{\frac{m}{2}})}
  \widetilde{R}^{\frac{m}{2}+1},
\\
  U_3
  &
  =
  \widetilde{R}^{\frac{3m}{2}}
  V^{\frac{m}{2}}
  \cdots
  \widetilde{R}^{m+2}
  V^2
  \widetilde{R}^{m+1}
  V^1.
\end{align*}
There are three cases of acceptance in the constructed proof system. In the first case, the verifier $W$
performs the \textsc{Rewinding Test} and accepts in Step~3.1. This happens with probability $\frac{p_1}{2}$,
where
\[
p_1 = \norm{\Pi_{\acc} V^{\frac{m}{2}+1} U_1 \ket{\psi}}^2.
\]
In the second case, the verifier $W$ performs the \textsc{Rewinding Test} and accepts in Step~3.5. This
happens with probability $\frac{p_2}{2}$, where
\[
p_2 = \norm{
  \Pi_{\acc} V^{\frac{m}{2}+1} U_3 Z U_2 \conjugate{(V^{\frac{m}{2}+1})} \Pi_{\rej} V^{\frac{m}{2}+1} U_1 \ket{\psi}}^2.
\]
Finally, in the third case, the verifier $W$ performs the \textsc{Invertibility Test} and accepts in
Step~4.3. This happens with probability $\frac{p_3}{2}$, where
\[
p_3 = \norm{\Pi_{\init} U_2 U_1 \ket{\psi}}^2.
\]
Hence, the total probability $p_{\acc}$ that $W$ accepts $x$ when communicating with ${R'_1, \ldots, R'_k}$
is given by ${p_{\acc} = \frac{1}{2} (p_1 + p_2 + p_3)}$.
From the soundness condition of the original proof system, it is obvious that ${p_1 \leq s}$. We shall show
that ${p_2 \leq 1 + 4 \sqrt{s} + 4s - p_3}$. This implies that ${p_{\acc} \leq \frac{1}{2} + 2 \sqrt{s} +
\frac{5s}{2}}$, and the soundness condition follows.

Using the triangle inequality, we have that
\begin{equation}
\begin{split}
& \norm{
  \Pi_{\acc} V^{\frac{m}{2}+1} U_3 Z U_2 \conjugate{(V^{\frac{m}{2}+1})} \Pi_{\rej} V^{\frac{m}{2}+1} U_1 \ket{\psi}
}
\\
& \hspace{1cm} \leq \norm{
  \Pi_{\acc} V^{\frac{m}{2}+1} U_3 Z U_2 \conjugate{(V^{\frac{m}{2}+1})} \Pi_{\rej} V^{\frac{m}{2}+1} U_1 \ket{\psi}
  -
  \Pi_{\acc} V^{\frac{m}{2}+1} U_3 Z U_2 U_1 \ket{\psi}
}
\\
& \hspace{15mm} + \norm{
  \Pi_{\acc} V^{\frac{m}{2}+1} U_3 Z U_2 U_1 \ket{\psi}
  -
  \Pi_{\acc} V^{\frac{m}{2}+1} U_3 Z \Pi_{\init} U_2 U_1 \ket{\psi}
}
\\
& \hspace{15mm} + \norm{\Pi_{\acc} V^{\frac{m}{2}+1} U_3 Z \Pi_{\init} U_2 U_1 \ket{\psi}}.
\end{split}
\label{Equation: bounding p_2}
\end{equation}
The first term of the right-hand side of inequality~(\ref{Equation: bounding p_2}) can be bounded from above
as follows:
\[
\begin{split}
& \norm{
  \Pi_{\acc} V^{\frac{m}{2}+1} U_3 Z U_2 \conjugate{(V^{\frac{m}{2}+1})} \Pi_{\rej} V^{\frac{m}{2}+1} U_1 \ket{\psi}
  -
  \Pi_{\acc} V^{\frac{m}{2}+1} U_3 Z U_2 U_1 \ket{\psi}
}
\\
& \hspace{1cm} \leq \norm{
  V^{\frac{m}{2}+1} U_3 Z U_2 \conjugate{(V^{\frac{m}{2}+1})} \Pi_{\rej} V^{\frac{m}{2}+1} U_1 \ket{\psi}
  -
  V^{\frac{m}{2}+1} U_3 Z U_2 U_1 \ket{\psi}
}
\\
& \hspace{1cm} = \norm{
  \conjugate{(V^{\frac{m}{2}+1})} \Pi_{\rej} V^{\frac{m}{2}+1} U_1 \ket{\psi}
  -
  U_1 \ket{\psi}
}
= \norm{
  \Pi_{\rej} V^{\frac{m}{2}+1} U_1 \ket{\psi}
  -
  V^{\frac{m}{2}+1} U_1 \ket{\psi}
}
\\
& \hspace{1cm} = \norm{
  - \Pi_{\acc} V^{\frac{m}{2}+1} U_1 \ket{\psi}
}
= \norm{
  \Pi_{\acc} V^{\frac{m}{2}+1} U_1 \ket{\psi}
} = \sqrt{p_1} \leq \sqrt{s}.
\end{split}
\]
The second term of the right-hand side of inequality~(\ref{Equation: bounding p_2}) can be bounded from above
as follows:
\[
\begin{split}
& \norm{
  \Pi_{\acc} V^{\frac{m}{2}+1} U_3 Z U_2 U_1 \ket{\psi}
  -
  \Pi_{\acc} V^{\frac{m}{2}+1} U_3 Z \Pi_{\init} U_2 U_1 \ket{\psi}
}
\\
& \hspace{1cm} \leq \norm{
  V^{\frac{m}{2}+1} U_3 Z U_2 U_1 \ket{\psi}
  -
  V^{\frac{m}{2}+1} U_3 Z \Pi_{\init} U_2 U_1 \ket{\psi}
}
\\
& \hspace{1cm} = \norm{
  U_2 U_1 \ket{\psi}
  -
  \Pi_{\init} U_2 U_1 \ket{\psi}
} = \norm{
  \Pi_{\illegal} U_2 U_1 \ket{\psi}
} = \sqrt{1 - p_3}.
\end{split}
\]
Here the last equality follows from the facts that ${
  U_2 U_1 \ket{\psi}
  =
  \Pi_{\init} U_2 U_1 \ket{\psi} + \Pi_{\illegal} U_2 U_1 \ket{\psi}
}$ is a unit vector, that ${\Pi_{\init} U_2 U_1 \ket{\psi}}$ and ${\Pi_{\illegal} U_2 U_1 \ket{\psi}}$ are
orthogonal, and that ${\norm{\Pi_{\init} U_2 U_1 \ket{\psi}}^2 = p_3}$.

Finally, since ${\Pi_{\init} U_2 U_1 \ket{\psi}}$ is an unnormalized state parallel to some legal initial
state and ${Z \Pi_{\init} = - \Pi_{\init}}$ from the definitions of $Z$ and $\Pi_{\init}$, the third term of
the right-hand side of inequality~(\ref{Equation: bounding p_2}) can be bounded as follows by using the
soundness condition of the original proof system:
\[
\norm{\Pi_{\acc} V^{\frac{m}{2}+1} U_3 Z \Pi_{\init} U_2 U_1 \ket{\psi}} = \norm{- \Pi_{\acc}
V^{\frac{m}{2}+1} U_3 \Pi_{\init} U_2 U_1 \ket{\psi}} = \norm{\Pi_{\acc} V^{\frac{m}{2}+1} U_3 \Pi_{\init}
U_2 U_1 \ket{\psi}} \leq \sqrt{s}.
\]
Putting everything together, we have
\[
\begin{split}
p_2 & = \norm{
  \Pi_{\acc} V^{\frac{m}{2}+1} U_3 Z U_2 \conjugate{(V^{\frac{m}{2}+1})} \Pi_{\rej} V^{\frac{m}{2}+1} U_1 \ket{\psi}}^2
\\
& \leq (2 \sqrt{s} + \sqrt{1 - p_3})^2 = 1 + 4 \sqrt{s (1 - p_3)} + 4s - p_3 \leq 1 + 4 \sqrt{s} + 4s - p_3,
\end{split}
\]
as desired.
\end{proof}

Now Theorem~\ref{Theorem: QMIP with perfect completeness} follows immediately from Lemma~\ref{Lemma: making
QMIP perfect complete} by appropriately applying sequential repetition.




\section{Parallelizing to Three Turns} \label{Subsection: Parallelizing to Three Turns}

In this section we prove Theorem~\ref{Theorem: parallelization to 3 turns}, which reduces the number of turns
to three without changing the number of provers. This is done by repeatedly converting any ${(2^l+1)}$-turn
QMIP system into a ${(2^{l-1}+1)}$-turn QMIP system where the gap decreases, but is still bounded by an
inverse-polynomial. We first show the following lemma.
\begin{lemma}
Let  $c^2>s$. Then, ${
  \QMIP(k, 4m + 1, c, s)
  \subseteq
  \QMIP \left( k, 2m + 1, \frac{1+c}{2}, \frac{1+\sqrt{s}}{2} \right)
}$. \label{Lemma: reducing number of turns by half}
\end{lemma}

\begin{proof}

Let $L$ be a language in ${\QMIP(k, 4m + 1,c, s)}$ and let $V$ be the corresponding ${(4m+1)}$-turn quantum
verifier. We construct a ${(2m+1)}$-turn quantum verifier $W$ for the new quantum $k$-prover interactive
proof system for $L$. The idea is that $W$ first receives the snapshot state that $V$ would have in
$(\sfV,\sfM_1,\ldots ,\sfM_k)$ just after the ${(2m+1)}$-st turn of the original system. $W$ then executes
with equal probability either a forward-simulation of the original system from the ${(2m+1)}$-st turn or a
backward-simulation of the original system from the ${(2m+1)}$-st turn. In the former case, $W$ accepts if
and only if the simulation results in acceptance in the original proof system, while in the latter case $W$
accepts if and only if the qubits in ${\sfV}$ are in state $\ket{0\cdots 0}$.\footnote{Recall that in the
original proof system the first turn was done by the provers, hence we do not measure the qubits in each
${\sfM}_i$ here.} The details are given in Figure~\ref{Figure: reducing the number of turns by half}.

\begin{figure}[ht]
\begin{algorithm*}{Verifier's Protocol to Reduce the Number of Turns by Half}
\begin{step}
\item
  Receive
  ${\sfV}$ and $\sfM_1$ from the first prover
  and ${\sfM}_i$ from the $i$th prover for ${2 \leq i \leq k}$.
\item
  Choose ${b \in \Binary}$ uniformly at random.
\item
  If ${b=0}$, execute a forward-simulation of the original proof system as follows:
  \begin{step}
  \item
    Apply ${V^{m+1}}$ to the qubits in
    ${({\sfV}, {\sfM}_1, \ldots, {\sfM}_k)}$.
    Send $b$ and ${\sfM}_i$ to the $i$th prover,
    for ${1 \leq i \leq k}$.
  \item
    For ${j = m + 2}$ to $2m$, do the following:\\
    Receive ${\sfM}_i$ from the $i$th prover,
    for ${1 \leq i \leq k}$.
    Apply $V^j$ to the qubits in ${({\sfV}, {\sfM}_1, \ldots, {\sfM}_k)}$.
    Send ${\sfM}_i$ to the $i$th prover, for ${1 \leq i \leq k}$.
  \item
    Receive ${\sfM}_i$ from the $i$th prover,
    for ${1 \leq i \leq k}$.
    Apply $V^{2m+1}$ to the qubits in
    ${({\sfV}, {\sfM}_1, \ldots, {\sfM}_k)}$.
    Accept if the content of
    ${({\sfV}, {\sfM}_1, \ldots, {\sfM}_k)}$
    is an accepting state of the original proof system,
    and reject otherwise.
  \end{step}
\item
  If ${b=1}$, execute a backward-simulation of the original proof system as follows:
  \begin{step}
  \item
    Send $b$ and ${\sfM}_i$ to the $i$th prover,
    for ${1 \leq i \leq k}$.
  \item
    For ${j = m}$ down to $2$, do the following:\\
    Receive ${\sfM}_i$ from the $i$th prover,
    for ${1 \leq i \leq k}$.
    Apply $\conjugate{(V^j)}$ to the qubits in ${({\sfV}, {\sfM}_1, \ldots, {\sfM}_k)}$.
    Send  ${\sfM}_i$ to the $i$th prover, for ${1 \leq i \leq k}$.
  \item
    Receive  ${\sfM}_i$ from the $i$th prover,
    for ${1 \leq i \leq k}$.
    Apply $\conjugate{(V^1)}$ to the qubits in ${({\sfV}, {\sfM}_1, \ldots, {\sfM}_k)}$.
    Accept if the qubits in ${\sfV}$ are in state $\ket{0\ldots 0}$,
    and reject otherwise.
  \end{step}
\end{step}
\end{algorithm*}
\caption{Verifier's protocol to reduce the number of turns by half.} \label{Figure: reducing the number of
turns by half}
\end{figure}

{\em Completeness:} Assume the input $x$ is in $L$. Let $P_1,\ldots ,P_k$ be the honest quantum provers in
the original proof system with a priori shared state $\ket{\Phi}$. Let $\ket{\psi_{2m+1}}$ be the quantum
state in ${({\sfV}, {\sfM}_1, \ldots, {\sfM}_k, {\sfP}_1, \ldots, {\sfP}_k)}$ just after the ${(2m+1)}$-st
turn in the original proof system. We construct honest provers $R_1,\ldots ,R_k$ for the new ${(2m +
1)}$-turn system. In addition to ${\sfV}$ and ${\sfM}_1$, $R_1$ prepares ${\sfP}_1$ in his private space.
Similarly, in addition to ${\sfM}_i$, $R_i$ prepares ${\sfP}_i$ in his private space for ${2 \leq i \leq k}$.
${R_1, \ldots, R_k}$ initially share $\ket{\psi_{2m+1}}$ in $(\sfV,\sfM_1,\ldots ,\sfM_k,\sfP_1,\ldots
,\sfP_k)$. At the first turn of the constructed proof system, $R_1$ sends ${\sfV}$ and ${\sfM}_1$ to $W$,
while each $R_i$,  for ${2 \leq i \leq k}$, sends ${\sfM}_i$ to $W$. At the $(2j-1)$-st turn for ${2 \leq j
\leq m + 1}$, if ${b=0}$, each $R_i$ applies $P_i^{m+j}$ (i.e. $P_i$'s transformation at the $(2m+2j-1)$-st
turn in the original system) while if ${b=1}$, each $R_i$ applies $\conjugate{(P_i^{m-j+3})}$ (i.e. the
inverse of $P_i$'s transformation at the $(2m-2j+5)$-th turn in the original system) to the qubits in
$(\sfP_i,\sfM_i)$, for $1 \leq i \leq k$. The provers ${R_1, \ldots, R_k}$ can then clearly convince $W$ with
probability at least $c$ if ${b=0}$, and with certainty if ${b=1}$. Hence, $W$ accepts every input ${x \in
L}$ with probability at least $\frac{1+c}{2}$.

{\em Soundness:} Now suppose that $x$ is not in $L$. Let $R'_1,\ldots ,R'_k$ be arbitrary provers for the
constructed proof system, and let $\ket{\psi}$ be an arbitrary quantum state that represents the state just
after the first turn in the constructed system. Suppose that, at the $(2j-1)$-st turn for ${2 \leq j \leq m +
1}$, each $R'_i$ applies $X_i^{j}$ if ${b=0}$ and $Y_i^{j}$ if ${b=1}$, for ${1 \leq i \leq k}$ and write
${\widetilde{X}^j = X_1^{j} \otimes \cdots \otimes X_k^{j}}$ and ${\widetilde{Y}^j = Y_1^{j} \otimes \cdots
\otimes Y_k^{j}}$. Define unitary transformations $U_0$ and $U_1$ by ${
  U_0
  =
  V^{2m+1}
  \widetilde{X}^{m+1}
  V^{2m}
  \cdots
  \widetilde{X}^2
  V^{m+1}
}$ and ${
  U_1
  =
  \conjugate{(V^1)}
  \widetilde{Y}^{m+1}
  \cdots
  \conjugate{(V^{m})}
  \widetilde{Y}^2
}$,
and let ${ \ket{\alpha} = \frac{1}{\norm{\Pi_{\acc} U_0 \ket{\psi}}} \Pi_{\acc} U_0 \ket{\psi} }$ and ${
\ket{\beta} = \frac{1}{\norm{\Pi_{\init} U_1 \ket{\psi}}} \Pi_{\init} U_1 \ket{\psi} }$, where $\Pi_{\acc}$
is the projection onto accepting states in the original proof system and $\Pi_{\init}$ is the projection on
$\ket{0\cdots 0}_{\sfV}$ in ${\sfV}$. Then
\[
\norm{\Pi_{\acc} U_0 \ket{\psi}} = \frac{1}{\norm{\Pi_{\acc} U_0 \ket{\psi}}} \bigabs{\bra{\psi}
\conjugate{U_0} \Pi_{\acc} U_0 \ket{\psi}} = F \bigl( \ketbra{\alpha}, U_0 \ketbra{\psi} \conjugate{U_0}
\bigr) = F \bigl( \conjugate{U_0} \ketbra{\alpha} U_0, \ketbra{\psi} \bigr),
\]
and thus, the probability $p_0$ of acceptance when ${b=0}$ is given by ${ p_0
= F \bigl( \conjugate{U_0} \ketbra{\alpha} U_0, \ketbra{\psi}\bigr)^2 }$. Similarly, the probability $p_1$ of
acceptance when ${b=1}$ is given by ${ p_1
= F \bigl( \conjugate{U_1} \ketbra{\beta} U_1, \ketbra{\psi} \bigr)^2 }$. Hence the probability $p_{\acc}$
that $W$ accepts $x$ when communicating with ${R'_1, \ldots, R'_k}$ is given by
\[
p_{\acc} = \frac{1}{2} (p_0 + p_1) = \frac{1}{2} \Bigl(
  F \bigl( \conjugate{U_0} \ketbra{\alpha} U_0, \ketbra{\psi} \bigr)^2
  +
  F \bigl( \conjugate{U_1} \ketbra{\beta} U_1, \ketbra{\psi} \bigr)^2
\Bigr).
\]
Therefore, from Lemma~\ref{Lemma: F(a,b)^2 + F(b,c)^2 < 1 + F(a,c)}, we have
\[
p_{\acc} \leq \frac{1}{2} \left(
  1
  +
  F \bigl(
      \conjugate{U_0} \ketbra{\alpha} U_0,
      \conjugate{U_1} \ketbra{\beta} U_1
    \bigr)
\right) = \frac{1}{2} \left(
  1
  +
  F \bigl(
      \ketbra{\alpha},
      U_0 \conjugate{U_1} \ketbra{\beta} U_1 \conjugate{U_0}
    \bigr)
\right).
\]
Note that ${\Pi_{\init} \ket{\beta} = \ket{\beta}}$ and that $\ket{\beta}$ is a legal quantum state which
could appear in the original proof system just after the first turn. Hence, from the soundness property of
the original proof system,
\[
\bignorm{\Pi_{\acc} U_0 \conjugate{U_1} \ket{\beta}}^2 = \bignorm{
  \Pi_{\acc}
  V^{2m+1}
  \widetilde{X}^{m+1}
  V^{2m}
  \cdots
  \widetilde{X}^2
  V^{m+1}
  \conjugate{(\widetilde{Y}^2)}
  V^{m}
  \cdots
  \conjugate{(\widetilde{Y}^{m+1})}
  V^1
  \ket{\beta}
}^2 \leq s,
\]
since ${ V^1, \conjugate{(\widetilde{Y}^{m+1})}, \cdots, V^{m}, \conjugate{(\widetilde{Y}^2)}, V^{m+1},
\widetilde{X}^2, \cdots, V^{2m}, \widetilde{X}^{m+1}, V^{2m+1} }$ form a legal sequence of transformations in
the original proof system.

Now, from the fact that ${\Pi_{\acc} \ket{\alpha} = \ket{\alpha}}$, we have
\[
F \bigl(
    \ketbra{\alpha},
    U_0 \conjugate{U_1} \ketbra{\beta} U_1 \conjugate{U_0}
  \bigr)
= \bigabs{\bra{\alpha} U_0 \conjugate{U_1} \ket{\beta}} = \bigabs{\bra{\alpha} \Pi_{\acc} U_0 \conjugate{U_1}
\ket{\beta}} \leq \norm{\Pi_{\acc} U_0 \conjugate{U_1} \ket{\beta}} \leq \sqrt{s}.
\]
Hence the probability $p_{\acc}$ that $W$ accepts $x$ is bounded by ${p_{\acc} \leq \frac{1}{2} +
\frac{\sqrt{s}}{2}}$, which completes the proof.
\end{proof}

Now, by repeatedly applying the construction in the proof of Lemma~\ref{Lemma: reducing number of turns by
half}, we can reduce the number of turns to three. The proof is straightforward, but we need to carefully
keep track of the efficiency of the constructed verifiers in each application, since the construction is
sequentially applied a logarithmic number of times.

\begin{lemma}
For any ${m \geq 4}$ and any $c,s$ such that $ \varepsilon=1-c$ and $\delta=1-s$ satisfy ${\delta
> 2 (m-1) \varepsilon}$,
 ${
  \QMIP(k, m, 1 - \varepsilon, 1 - \delta)
  \subseteq
  \QMIP \left( k, 3, 1 - \frac{2 \varepsilon}{m-1}, 1 - \frac{\delta}{(m-1)^2} \right)
}$. \label{Theorem: reducing number of turns to three}
\end{lemma}

\begin{proof}
Let $l$ be such that ${2^l + 1 \leq m \leq 2^{l+1} + 1}$. Trivially, ${ \QMIP(k, m, c, s) \subseteq \QMIP(k,
2^{l+1}+1, c, s) }$. We show $\QMIP(k, 2^{l+1}+1, 1-\varepsilon, 1-\delta)  \subseteq \QMIP ( k, 3, 1 -
\frac{2 \varepsilon}{m-1}, 1 - \frac{\delta}{(m-1)^2} )$.

Let $L$ be a language in ${\QMIP(k, 2^{l+1}+1, 1 - \varepsilon, 1 - \delta)}$ and let $V^{(0)}$ be the
corresponding ${(2^{l+1} + 1)}$-turn quantum verifier. Given a description of $V^{(0)}$ one can compute in
polynomial time  a description of a ${(2^l + 1)}$-turn quantum verifier $V^{(1)}$ following the proof of
Lemma~\ref{Lemma: reducing number of turns by half}. The resulting proof system has completeness at least ${1
- \frac{\varepsilon}{2}}$ and soundness at most ${\frac{1}{2} + \frac{\sqrt{1 - \delta}}{2} \leq 1 -
\frac{\delta}{4}}$. Crucially, the description of $V^{(1)}$ is at most some constant times the size of the
description of ${V^{(0)}}$ plus an amount bounded by a polynomial in $\abs{x}$. Hence it is obvious that,
given a description of ${V^{(0)}}$, one can compute in polynomial time  a description of a three-turn quantum
verifier $V^{(l)}$ by repeatedly applying the construction in the proof of Lemma~\ref{Lemma: reducing number
of turns by half} $l$ times. The resulting proof system has completeness at least ${1 -
\frac{\varepsilon}{2^l} \geq 1 - \frac{2\varepsilon}{m-1}}$ and soundness at most ${1 - \frac{\delta}{4^l}
\leq 1 - \frac{\delta}{(m-1)^2}}$, as desired.
\end{proof}

Theorem \ref{Theorem: parallelization to 3 turns} now follows immediately from Theorem~\ref{Theorem: QMIP
with perfect completeness}~and Lemma~\ref{Theorem: reducing number of turns to three}: For every $p \in
\poly$ there is an $m' \in \poly$ such that ${ \QMIP(k, m, c, s) \subseteq \QMIP(k, m', 1, 2^{-p}) }$
$\subseteq \QMIP \left( k, 3, 1, 1 - \frac{1 - 2^{-p}}{(m'-1)^2} \right)$. Now it suffices to observe that
$\frac{1 - 2^{-p}}{(m'-1)^2} \in \poly^{-1}$.


\section{Public-Coin Systems} \label{Section:part3}

In this section we present the last part to complete the proof of Theorem \ref{Theorem: parallelization to 1
round}. We show how any three-turn  QMIP system with sufficiently large gap can be converted into a two-turn
QMIP system with one extra prover, in which the gap is bounded by an inverse-polynomial. Although we also
have a direct proof for this, given in Appendix~\ref{Appendix: Direct Proof of Modifing Three-Turn Systems to
Two-Turn Systems}, we will take a detour by showing how (i) any three-turn  QMIP system with sufficiently
large gap can be modified to a three-turn \emph{public-coin} QMIP system with inverse-polynomial gap without
changing the number of provers, and (ii) any three-turn public-coin QMIP system can be converted into a
two-turn QMIP system without changing completeness and soundness, by adding an extra prover. The added
benefits of our detour are a proof of the equivalence of public-coin QMIP systems and general QMIP systems
(Theorem~\ref{Theorem: public-coin QMIP = general QMIP}) and a proof that $\QIP$ and hence $\PSPACE$ has a
two-prover one-round quantum interactive proof system of perfect completeness and exponentially small
soundness (Corollary~\ref{Corollary: QIP is in QMIP(2,2,1,s)}).\footnote{The direct proof in
Appendix~\ref{Appendix: Direct Proof of Modifing Three-Turn Systems to Two-Turn Systems} would only give the
weaker corollary that $\QIP$ has a two-prover one-round quantum interactive proof system of perfect
completeness, but with soundness only exponentially close to $\frac{1}{2}$. This is indeed weaker than what
we can show with the detour, since it is not known how to amplify the success probability of QMIP systems
without increasing either the number of provers or the number of turns.}

\subsection{Converting to Public-Coin Systems}

In this subsection we prove Theorem~\ref{Theorem: public-coin QMIP = general QMIP} showing that any language
that has a quantum $k$-prover interactive proof system with two-sided bounded error also has a {\em
public-coin} quantum $k$-prover interactive proof system of perfect completeness and exponentially small
soundness.

We first show that any three-turn QMIP system with sufficiently large gap can be modified to a three-turn
public-coin QMIP system with the same number of provers and inverse-polynomial gap. In the single-prover
case, Marriott~and~Watrous~\cite{MarWat05CC} proved a similar statement. Our proof is a generalization of
their proof (Theorem~5.4~in~Ref.~\cite{MarWat05CC}) to the multi-prover case.

\begin{lemma}
For any $c,s$ satisfying $c^2 > s$, $\QMIP(k,3,c,s) \subseteq
\publicQMIP(k,3,\frac{1+c}{2},\frac{1+\sqrt{s}}{2})$. Moreover, the message from the verifier to each prover
in the public-coin system consists of only one classical bit. \label{Theorem: making 3-turn systems
public-coin}
\end{lemma}

\begin{proof}
Let $L$ be a language in ${\QMIP(k, 3, c, s)}$ and let $V$ be the corresponding three-turn quantum verifier.
We construct a new verifier $W$ for the public-coin system. The idea is that in the first turn $W$ receives
the reduced state in the original register ${\sfV}$ of the snapshot state just after the second turn (i.e.,
just after the first transformation of $V$) in the original proof system. $W$ then flips a fair classical
coin $b \in \{0,1\}$ and broadcasts $b$ to the provers. At the third turn the $i$th prover is requested to
send the register $\sfM_i$ of the original proof system, for $1 \leq i \leq k$. If $b=0$ the qubits in
${({\sfV}, {\sfM}_1, \ldots, {\sfM}_k)}$ should form the quantum state the original verifier $V$ would
possess just after the third turn of the original proof system. Now $W$ applies $V^2$ to the qubits in
${({\sfV}, {\sfM}_1, \ldots, {\sfM}_k)}$ and accepts if and only if the content of ${({\sfV}, {\sfM}_1,
\ldots, {\sfM}_k)}$ is an accepting state of the original proof system. On the other hand, if ${b=1}$, the
qubits in ${({\sfV}, {\sfM}_1, \ldots, {\sfM}_k)}$ should form the quantum state the original verifier $V$
would possess just after the second turn of the original proof system. Now $W$ applies $\conjugate{(V^1)}$ to
the qubits in ${({\sfV}, {\sfM}_1, \ldots, {\sfM}_k)}$ and accepts if and only if all the qubits in ${\sfV}$
are in state $\ket{0}$. The detailed description of the protocol of $W$ is given in Figure~\ref{Figure:
verifier's protocol in three-turn public-coin system}. The analysis of completeness and soundness of the
constructed proof system is nearly identical to the one in Lemma \ref{Lemma: reducing number of turns by
half}, and is relegated to Appendix \ref{Appendix: proof of theorem for making 3-turn systems public-coin}.
\end{proof}

\begin{figure}[h!]
\begin{algorithm*}{Verifier's Protocol in Three-Turn Public-Coin System}
\begin{step}
\item
  Receive ${\sfV}$ from the first prover
  and receive nothing from the $i$th prover, for ${2 \leq i \leq k}$.
\item
  Choose ${b \in \Binary}$ uniformly at random.
  Send $b$ to each prover.
\item
  Receive ${\sfM}_i$
  from the $i$th prover for ${1 \leq i \leq k}$.
  \begin{step}
  \item
    If ${b=0}$,
    apply $V^2$ to the qubits in ${({\sfV}, {\sfM}_1, \ldots, {\sfM}_k)}$.
    Accept if the content of ${({\sfV}, {\sfM}_1, \ldots, {\sfM}_k)}$
    is an accepting state of the original proof system,
    and reject otherwise.
  \item
    If ${b=1}$,
    apply $\conjugate{(V^1)}$ to the qubits in ${({\sfV}, {\sfM}_1, \ldots, {\sfM}_k)}$.
    Accept if all the qubits in ${\sfV}$ are in state $\ket{0}$,
    and reject otherwise.
  \end{step}
\end{step}
\end{algorithm*}
\caption{Verifier's protocol in three-turn public-coin system.} \label{Figure: verifier's protocol in
three-turn public-coin system}
\end{figure}

Theorem \ref{Theorem: public-coin QMIP = general QMIP} now follows directly from Theorem~\ref{Theorem:
parallelization to 3 turns}~and~Lemma \ref{Theorem: making 3-turn systems public-coin} together with
sequential repetition: Theorem~\ref{Theorem: parallelization to 3 turns}~and~Lemma \ref{Theorem: making
3-turn systems public-coin} imply that there is a $p' \in \poly$ such that ${
  \QMIP(k, m, c, s)
  \subseteq
  \QMIP \bigl( k, 3, 1, 1 - \frac{1}{p'} \bigr)
  \subseteq
  \publicQMIP \bigl(  k, 3, 1, 1 - \frac{1}{4p'} \bigr)
}$, since ${
  \frac{1}{2} \left( 1 +\sqrt{1 - \frac{1}{p'}} \right)
  \leq
  1 - \frac{1}{4p'}
}$. Finally, sequential repetition gives that for all ${p \in \poly}$ there exists an ${m' \in \poly}$ such
that ${
  \publicQMIP \bigl(  k, 3, 1, 1 - \frac{1}{4p'} \bigr)
  \subseteq
  \publicQMIP(k, m', 1, 2^{-p})
}$.


\subsection{Parallelizing to Two Turns}
\label{Subsection: Parallelizing to Two Turns}

Finally, we prove the last piece of Theorem~\ref{Theorem: parallelization to 1 round} by showing that any
three-turn public-coin  quantum $k$-prover interactive proof system can be converted into a two-turn (i.e.,
one-round) ${(k+1)}$-prover system without changing completeness and soundness. The idea of the proof is to
send questions only to the first $k$ provers to request the original second messages from the $k$ provers in
the original system and to receive from the ${(k+1)}$-st prover the original first messages of the $k$
provers in the original system without asking him any question.

\begin{lemma}
${\publicQMIP(k,3,c,s)\subseteq \QMIP(k+1, 2, c, s)}$. \label{Theorem: reducing number of turns to two}
\end{lemma}

\begin{proof}
Let $L$ be a language in $\publicQMIP(k,m,c,s)$ and let $V$ be the corresponding verifier.

The protocol can be viewed as follows: At the first turn, $V$ first receives a quantum register $\sfM_i$ from
the $i$th prover, for each ${1 \leq i \leq k}$. $V$ flips a fair classical coin $q_{\sfM}$ times to generate
a random string $r$ of length $q_{\sfM}$, and broadcasts $r$ to all the provers. $V$ also stores $r$ in a
quantum register $\sfQ$ in his private space. Finally, at the third turn, $V$ receives a quantum register
$\sfN_i$ from the $i$th prover, for each ${1 \leq i \leq k}$. $V$ then prepares a quantum register $\sfV$ for
his work space, where all the qubits in $\sfV$ are initialized to state $\ket{0}$, applies the transformation
$V^{\final}$ to the qubits in ${(\sfQ, \sfV, \sfM_1, \ldots, \sfM_k, \sfN_1, \ldots, \sfN_k)}$, and performs
the measurement ${\Pi = \{ \Pi_{\acc}, \Pi_{\rej} \}}$ to decide acceptance or rejection. We construct a
two-turn quantum verifier $W$ for the new quantum ${(k+1)}$-prover interactive proof system for $L$.

The constructed prover $W$ starts with generating a random string $r$ of length $q_{\sfM}$ in the first turn,
and sends $r$ to the first $k$ provers. $W$ does not send any question to the last prover. In the second turn
$W$ receives $\sfN_i$ from the $i$th prover expecting the original second message from the original $i$th
prover, for $1 \leq i \leq k$. From the ${(k+1)}$-st prover $W$ receives $k$ quantum registers ${\sfM_1,
\ldots, \sfM_k}$, expecting the original first messages of the original $k$ provers. $W$ then proceeds like
$V$ would. A detailed description of the protocol of $W$ is given in Figure~\ref{Figure: verifier's protocol
in one-round system}.

\begin{figure}[h!]
\begin{algorithm*}{Verifier's Protocol in One-Round System}
\begin{step}
\item
  Prepare a quantum register ${\sfV}$,
  and initialize all the qubits in ${\sfV}$ to state $\ket{0}$.
  Flip a fair classical coin $q_{\sfM}$ times
  to generate a random string $r$ of length $q_{\sfM}$.
  Store $r$ in a quantum register ${\sfQ}$,
  and send $r$ to the $i$th prover for ${1 \leq i \leq k}$.
  Send nothing to the ${(k+1)}$-st prover.
\item
 Receive a quantum register $\sfN_i$ from the $i$th prover,
  for ${1 \leq i \leq k}$,
  and $k$ quantum registers ${\sfM_1, \ldots, \sfM_k}$
  from the ${(k+1)}$-st prover.
  Apply $V^{\final}$
  to the qubits in ${(\sfQ, \sfV, \sfM_1, \ldots, \sfM_k, \sfN_1, \ldots, \sfN_k)}$
  and accept if and only if
  the content of ${(\sfQ, \sfV, \sfM_1, \ldots, \sfM_k, \sfN_1, \ldots, \sfN_k)}$
  is an accepting state of the original proof system.
\end{step}
\end{algorithm*}
\caption{Verifier's protocol to reduce the number of turns to two.} \label{Figure: verifier's protocol in
one-round system}
\end{figure}

{\em Completeness:} Assume  the input $x$ is in $L$. Let $P_1,\ldots ,P_k$ be the honest provers in the
original proof system. Let $\ket{\psi_1}$ be the quantum state in ${({\sfM}_1, \ldots, {\sfM}_k, {\sfP}_1,
\ldots, {\sfP}_k)}$ in the original proof system just after the first turn. We construct honest provers
$R_1,\ldots ,R_{k+1}$ for the two-turn system. For ${1 \leq i \leq k}$, $R_i$ prepares quantum register
${\sfP}_i$ in his private space, where some of the qubits in $\sfP_i$ form the quantum register $\sfN_i$,
while $R_{k+1}$ prepares the quantum registers ${{\sfM}_1, \ldots, {\sfM}_k}$ in his private space. ${R_1,
\ldots, R_{k+1}}$ initially share $\ket{\psi_1}$ in ${(\sfM_1, \ldots, \sfM_k, \sfP_1, \ldots, \sfP_k)}$. At
the second turn, $R_{k+1}$ just sends the qubits in ${({\sfM}_1, \ldots, {\sfM}_k)}$ to $W$, while each
$R_i$, after receiving $r$, just behaves like $P_i$ would at the third turn of the original system, and then
sends ${\sfN}_i$, which is a part of $\sfP_i$, to $W$, for ${1 \leq i \leq k}$. It is obvious from the
construction that the provers ${R_1, \ldots, R_{k+1}}$ can convince $W$ with the same probability with which
${P_1, \ldots, P_k}$ could convince $V$, which is at least $c$.

{\em Soundness:} Now assume the input $x$ is not in $L$. Let ${R'_1, \ldots, R'_{k+1}}$ be any provers for
the constructed proof system and let $\sfR'_i$ be the quantum register consisting of all the qubits in the
private space of $R'_i$, for ${1 \leq i \leq k+1}$. For $R'_{k+1}$, some of the qubits in $\sfR'_{k+1}$ form
the register ${\sfM = (\sfM_1, \ldots, \sfM_k)}$. Let $\ket{\psi}$ be an arbitrary quantum state in
${(\sfR'_1, \ldots, \sfR'_{k+1})}$ that is initially shared by ${R'_1, \ldots, R'_{k+1}}$. Suppose that, at
the second turn, each $R'_i$ applies $X_i^{(r)}$, for ${1 \leq i \leq k}$, if the message from $W$ is $r$.
Without loss of generality, we assume that $R'_{k+1}$ does nothing, and just sends the qubits in ${(\sfM_1,
\ldots, \sfM_k)}$ at the second turn, since $R'_{k+1}$ receives nothing from $W$ (that $R'_{k+1}$ applies
some transformation $Z$ is equivalent to sharing ${Z \ket{\psi}}$ at the beginning).

Consider three-turn quantum provers ${P'_1, \ldots, P'_k}$ for the original proof system with the following
properties: (1) each $P'_i$ prepares the quantum register ${\sfR}'_i$ in his private space, for ${1 \leq i
\leq k}$, (2) ${P'_1, \ldots, P'_k}$ initially share $\ket{\psi}$ in ${({\sfR}'_1, \ldots, {\sfR}'_{k+1})}$,
where all the qubits in ${\sfR}'_{k+1}$ except for those in ${({\sfM}_1, \ldots, {\sfM}_k)}$ are shared
arbitrarily, (3) at the first turn, each $P'_i$ sends ${\sfM}_i$ to $V$, for ${1 \leq i \leq k}$, and (4) if
the message from $V$ is $r$, at the third turn, each $P'_i$ applies $X_i^{(r)}$ to the qubits in $\sfR'_i$,
for ${1 \leq i \leq k}$. It is obvious that these provers ${P'_1, \ldots, P'_k}$ can convince the original
verifier $V$ with the same probability that ${R'_1, \ldots, R'_{k+1}}$ can convince $W$. Hence, the
probability $W$ accepts $x$ is at most $s$, as desired.
\end{proof}

Now Theorem~\ref{Theorem: parallelization to 1 round} follows from Theorem~\ref{Theorem: parallelization to 3
turns} and Lemmas~\ref{Theorem: making 3-turn systems public-coin}~and~\ref{Theorem: reducing number of turns
to two}.
Corollary~\ref{Corollary: QIP is in QMIP(2,2,1,s)}, claiming ${\QIP \subseteq \QMIP(2, 2, 1, 2^{-p})}$ for
any ${p \in \poly}$ follows directly from Lemma~\ref{Theorem: reducing number of turns to two} and the fact
shown by Marriott~and~Watrous~\cite{MarWat05CC} that any language in $\QIP$ can be verified by a
three-message public-coin quantum interactive proof system of perfect completeness with exponentially small
error in soundness (i.e., ${\QIP \subseteq \QMAM(1, 2^{-p})}$ for any ${p \in \poly}$).



\begin{thebibliography}{BOGKW88}
\expandafter\ifx\csname urlstyle\endcsname\relax
  \providecommand{\doi}[1]{doi:\discretionary{}{}{}#1}\else
  \providecommand{\doi}{doi:\discretionary{}{}{}\begingroup
  \urlstyle{rm}\Url}\fi

\bibitem[AKN98]{AhaKitNis98STOC}
D.~Aharonov, A.~Yu. Kitaev, and N.~Nisan.
\newblock Quantum circuits with mixed states.
\newblock In \emph{Proceedings of the Thirtieth Annual ACM Symposium on Theory
  of Computing}, pages 20--30. 1998.

\bibitem[ALM{\etalchar{+}}98]{AroLunMotSudSze98JACM}
S.~Arora, C.~Lund, R.~Motwani, M.~Sudan, and M.~Szegedy.
\newblock Proof verification and the hardness of approximation problems.
\newblock \emph{Journal of the ACM}, 45(3):501--555, 1998.

\bibitem[AS98]{AroSaf98JACM}
S.~Arora and S.~Safra.
\newblock Probabilistic checking of proofs: A new characterization of {$\NP$}.
\newblock \emph{Journal of the ACM}, 45(1):70--122, 1998.

\bibitem[Bab85]{Bab85STOC}
L.~Babai.
\newblock Trading group theory for randomness.
\newblock In \emph{Proceedings of the Seventeenth Annual ACM Symposium on
  Theory of Computing}, pages 421--429. 1985.

\bibitem[BFL91]{BabForLun91CC}
L.~Babai, L.~J. Fortnow, and C.~Lund.
\newblock Non-deterministic exponential time has two-prover interactive
  protocols.
\newblock \emph{Computational Complexity}, 1(1):3--40, 1991.

\bibitem[BOGKW88]{BenGolKilWig88STOC}
M.~Ben-Or, S.~Goldwasser, J.~Kilian, and A.~Wigderson.
\newblock Multi-prover interactive proofs: How to remove intractability
  assumptions.
\newblock In \emph{Proceedings of the Twentieth Annual ACM Symposium on Theory
  of Computing}, pages 113--131. 1988.

\bibitem[CCL94]{CaiConLip94JCSS}
J.-Y. Cai, A.~Condon, and R.~J. Lipton.
\newblock {$\PSPACE$} is provable by two provers in one round.
\newblock \emph{Journal of Computer and System Sciences}, 48(1):183--193, 1994.

\bibitem[CGJ07]{CleGavJai07arXiv}
R.~E. Cleve, D.~Gavinsky, and R.~Jain.
\newblock Entanglement-resistant two-prover interactive proof systems and
  non-adaptive private information retrieval systems, July 2007.
\newblock ArXiv.org e-Print archive, arXiv:0707.1729 [quant-ph].

\bibitem[CHTW04]{CleHoyTonWat04CCC}
R.~E. Cleve, P.~H{\o}yer, B.~F. Toner, and J.~H. Watrous.
\newblock Consequences and limits of nonlocal strategies.
\newblock In \emph{Nineteenth Annual {IEEE} Conference on Computational
  Complexity}, pages 236--249. 2004.

\bibitem[FGL{\etalchar{+}}96]{FeiGolLovSafSze96JACM}
U.~Feige, S.~Goldwasser, L.~Lov\'asz, S.~Safra, and M.~Szegedy.
\newblock Interactive proofs and the hardness of approximating cliques.
\newblock \emph{Journal of the ACM}, 43(2):268--292, 1996.

\bibitem[FL92]{FeiLov92STOC}
U.~Feige and L.~Lov\'asz.
\newblock Two-prover one-round proof systems: Their power and their problems
  (extended abstract).
\newblock In \emph{Proceedings of the Twenty-Fourth Annual ACM Symposium on the
  Theory of Computing}, pages 733--744. 1992.

\bibitem[FRS94]{ForRomSip94TCS}
L.~J. Fortnow, J.~Rompel, and M.~Sipser.
\newblock On the power of multi-prover interactive protocols.
\newblock \emph{Theoretical Computer Science}, 134(2):545--557, 1994.

\bibitem[GMR89]{GolMicRac89SIComp}
S.~Goldwasser, S.~Micali, and C.~W. Rackoff.
\newblock The knowledge complexity of interactive proof systems.
\newblock \emph{SIAM Journal on Computing}, 18(1):186--208, 1989.

\bibitem[IKP{\etalchar{+}}07]{yao:tsirelson}
T.~Ito, H.~Kobayashi, D.~Preda, X.~Sun, and A.~C.-C. Yao.
\newblock Generalized {Tsirelson} inequalities, commuting-operator provers, and
  multi-prover interactive proof systems, 2007.
\newblock Talk given at QIP'08, December 2007, Delhi, India.

\bibitem[KKM{\etalchar{+}}07]{KemKobMatTonVid07arXiv}
J.~Kempe, H.~Kobayashi, K.~Matsumoto, B.~F. Toner, and T.~Vidick.
\newblock Entangled games are hard to approximate, 2007.
\newblock ArXiv.org e-Print archive, arXiv:0704.2903 [quant-ph].

\bibitem[KM03]{KobMat03JCSS}
H.~Kobayashi and K.~Matsumoto.
\newblock Quantum multi-prover interactive proof systems with limited prior
  entanglement.
\newblock \emph{Journal of Computer and System Sciences}, 66(3):429--450, 2003.

\bibitem[KSV02]{KitSheVya02Book}
A.~Yu. Kitaev, A.~H. Shen, and M.~N. Vyalyi.
\newblock \emph{Classical and Quantum Computation}, volume~47 of \emph{Graduate
  Studies in Mathematics}.
\newblock American Mathematical Society, 2002.

\bibitem[KW00]{KitWat00STOC}
A.~Yu. Kitaev and J.~H. Watrous.
\newblock Parallelization, amplification, and exponential time simulation of
  quantum interactive proof systems.
\newblock In \emph{Proceedings of the Thirty-Second Annual ACM Symposium on
  Theory of Computing}, pages 608--617. 2000.

\bibitem[MW05]{MarWat05CC}
C.~Marriott and J.~H. Watrous.
\newblock Quantum {Arthur-Merlin} games.
\newblock \emph{Computational Complexity}, 14(2):122--152, 2005.

\bibitem[NC00]{NieChu00Book}
M.~A. Nielsen and I.~L. Chuang.
\newblock \emph{Quantum Computation and Quantum Information}.
\newblock Cambridge University Press, 2000.

\bibitem[NS03]{NaySho03PRA}
A.~Nayak and P.~W. Shor.
\newblock Bit-commitment-based quantum coin flipping.
\newblock \emph{Physical Review A}, 67(1):012304, 2003.

\bibitem[Sho96]{Sho96FOCS}
P.~W. Shor.
\newblock Fault-tolerant quantum computation.
\newblock In \emph{37th Annual Symposium on Foundations of Computer Science},
  pages 56--65. 1996.

\bibitem[SR02]{SpeRud02PRA}
R.~W. Spekkens and T.~Rudolph.
\newblock Degrees of concealment and bindingness in quantum bit-commitment
  protocols.
\newblock \emph{Physical Review A}, 65(1):012310, 2002.

\bibitem[Wat03]{Wat03TCS}
J.~H. Watrous.
\newblock {$\PSPACE$} has constant-round quantum interactive proof systems.
\newblock \emph{Theoretical Computer Science}, 292(3):575--588, 2003.

\bibitem[Wat06]{Wat06STOC}
J.~H. Watrous.
\newblock Zero-knowledge against quantum attacks.
\newblock In \emph{Proceedings of the 38th Annual ACM Symposium on Theory of
  Computing}, pages 296--305. 2006.

\bibitem[Weh06]{Weh06STACS}
S.~Wehner.
\newblock Entanglement in interactive proof systems with binary answers.
\newblock In \emph{STACS 2006, 23rd Annual Symposium on Theoretical Aspects of
  Computer Science}, volume 3884 of \emph{Lecture Notes in Computer Science},
  pages 162--171. 2006.

\end{thebibliography}

\newcommand{\etalchar}[1]{$^{#1}$}


\appendix


\section*{\appendixname}



\section{Proof of Lemma~\ref{Theorem: making 3-turn systems public-coin}}
\label{Appendix: proof of theorem for making 3-turn systems public-coin}

\begin{proof}
{\em Completeness:} Assume the input $x$ is in $L$. Let $P_1,\ldots ,P_k$ be the honest quantum provers in
the original proof system with a priori shared state $\ket{\Phi}$ in $(\sfP_1,\ldots,\sfP_k)$. Let
$\ket{\psi_{2}}$ be the quantum state in ${({\sfV}, {\sfM}_1, \ldots, {\sfM}_k, {\sfP}_1, \ldots, {\sfP}_k)}$
just after the second turn in the original proof system. We construct honest provers $R_1,\ldots ,R_k$ for
the public-coin system. In addition to ${\sfV}$ and ${\sfM_1}$, $R_1$ prepares ${\sfP}_1$ in his private
space. Similarly, in addition to ${\sfM}_i$, $R_i$ prepares ${\sfP}_i$ in his private space, for ${2 \leq i
\leq k}$. ${R_1, \ldots, R_k}$ initially share $\ket{\psi_{2}}$ in ${(\sfV, \sfM_1, \ldots, \sfM_k, \sfP_1,
\ldots, \sfP_k)}$. At the first turn of the constructed proof system, $R_1$ sends ${\sfV}$ to $W$, while each
$R_i$, ${2 \leq i \leq k}$ send nothing to $W$. At the third turn, if ${b=0}$ each $R_i$ applies $P_i^{2}$ to
the qubits in $(\sfM_i,\sfP_i)$ and then sends ${\sfM}_i$ to $W$, while if ${b=1}$, each $R_i$ does nothing
and sends ${\sfM}_i$ to $W$. It is obvious that the provers ${R_1, \ldots, R_k}$ can convince $W$ with
probability at least $c$ if ${b=0}$, and with certainty if ${b=1}$. Hence, $W$ accepts every input ${x \in
L}$ with probability at least $\frac{1 +c}{2}$.

{\em Soundness:} Now suppose  that $x$ is not in $L$. Let $R'_1,\ldots ,R'_k$ be arbitrary provers for the
constructed proof system, and let $\ket{\psi}$ be an arbitrary quantum state that represents the state just
after the first turn in the constructed system. Suppose that at the third turn each $R'_i$ applies $X_i$ if
${b=0}$ and $Y_i$ if ${b=1}$, for ${1 \leq i \leq k}$ and write ${\widetilde{X} = X_1 \otimes \cdots \otimes
X_k}$ and ${\widetilde{Y} = Y_1 \otimes \cdots \otimes Y_k}$. Note that $\widetilde{X}$ and $\widetilde{Y}$
are unitary transformations that do not act over the qubits in ${\sfV}$. Let ${ \ket{\alpha} =
\frac{1}{\norm{\Pi_{\acc} V^2 \widetilde{X} \ket{\psi}}} \Pi_{\acc} V^2 \widetilde{X} \ket{\psi} }$ and ${
\ket{\beta} = \frac{1}{\norm{\Pi_{\init} \conjugate{(V^1)} \widetilde{Y} \ket{\psi}}} \Pi_{\init}
\conjugate{(V^1)} \widetilde{Y} \ket{\psi} }$, where $\Pi_{\acc}$ is the projection onto accepting states in
the original proof system and $\Pi_{\init}$ is the projection onto states in which all the qubits in ${\sfV}$
are in state $\ket{0}$.

Then, with a similar argument to that in the proof of Lemma~\ref{Lemma: reducing number of turns by half},
the probability $p_{\acc}$ that $W$ accepts $x$ when communicating with ${R'_1, \ldots, R'_{k+1}}$ is bounded
by
\begin{align*}
p_{\acc} &\leq \frac{1}{2} \left(
  1
  +
  F \bigl(
      \conjugate{\widetilde{X}} \conjugate{(V^2)}
      \ketbra{\alpha}
      V^2 \widetilde{X},
      \conjugate{\widetilde{Y}} V^1
      \ketbra{\beta}
      \conjugate{(V^1)} \widetilde{Y}
    \bigr)
\right) \\
&= \frac{1}{2} \left(
  1
  +
  F \bigl(
      \ketbra{\alpha},
      V^2 \widetilde{X} \conjugate{\widetilde{Y}} V^1
      \ketbra{\beta}
      \conjugate{(V^1)} \widetilde{Y} \conjugate{\widetilde{X}} \conjugate{(V^2)}
    \bigr)
\right).
\end{align*}

Since ${\Pi_{\init} \ket{\beta} = \ket{\beta}}$ is a legal quantum state which could appear just after the
first turn in the original proof system, ${ V^1, \bigl( \widetilde{X} \conjugate{\widetilde{Y}} \bigr), V^2
}$ form a legal sequence of transformations in the original proof system, and ${\Pi_{\acc} \ket{\alpha} =
\ket{\alpha}}$, again a similar argument to that in the proof of Lemma~\ref{Lemma: reducing number of turns
by half} shows that ${ F \bigl(
    \ketbra{\alpha},
    V^2 \widetilde{X} \conjugate{\widetilde{Y}} V^1
    \ketbra{\beta}
    \conjugate{(V^1)} \widetilde{Y} \conjugate{\widetilde{X}} \conjugate{(V^2)}
  \bigr)
\leq \sqrt{s} }$.

Hence the probability $p_{\acc}$ that $W$ accepts $x$ is bounded by ${p_{\acc} \leq \frac{1}{2} +
\frac{\sqrt{s}}{2}}$, as desired.
\end{proof}


\section{Direct Proof of Modifying Three-Turn Systems to Two-Turn Systems}
\label{Appendix: Direct Proof of Modifing Three-Turn Systems to Two-Turn Systems}

For completeness, here we give a direct proof of the fact that any $k$-prover three-turn system can be
converted into a ${(k+1)}$-prover two-turn system.

\begin{theorem}
For any $c,s$ satisfying $c^2>{s}$, ${
  \QMIP(k, 3, c, s)
  \subseteq
  \QMIP \left( k+1, 2,  \frac{1+c}{2}, \frac{1+\sqrt{s}}{2} \right)
}$. \label{Theorem: reducing number of turns to two directly}
\end{theorem}

\begin{proof}
The proof is very similar to that of Lemma~\ref{Theorem: making 3-turn systems public-coin}. Indeed,  our
starting point is the same, but this time we move to a two-turn proof system, instead of a three-turn
public-coin system, by adding an extra prover. As in Lemma~\ref{Theorem: reducing number of turns to two}, we
first broadcast a random bit $b\in\{0,1\}$ to all but the extra prover, and ask the extra prover to send us a
register ${\sfV}$ and the other provers to send us registers ${\sfM_i}$. We then proceed as in Step 3 of the
proof system given in Lemma~\ref{Theorem: making 3-turn systems public-coin}: a detailed description is given
in Figure~\ref{Figure: verifier's protocol in one-round system (direct construction)}.

\begin{figure}[t!]
\begin{algorithm*}{Verifier's Protocol in One-Round System (Direct Construction)}
\begin{step}
\item
  Choose ${b \in \Binary}$ uniformly at random.
  Send $b$ only to the first $k$ provers,
  and send nothing to the ${(k+1)}$-st prover.
\item
  Receive  ${\sfM}_i$ from the $i$th prover,
  for ${1 \leq i \leq k}$,
  and  ${\sfV}$ from the ${(k+1)}$-st prover.
  \begin{step}
  \item
    If ${b=0}$,
    apply $V^2$ to the qubits in ${({\sfV}, {\sfM}_1, \ldots, {\sfM}_k)}$.
    Accept if the content of ${({\sfV}, {\sfM}_1, \ldots, {\sfM}_k)}$
    is an accepting state in the original proof system,
    and reject otherwise.
  \item
    If ${b=1}$,
    apply $\conjugate{(V^1)}$ to the qubits in ${({\sfV}, {\sfM}_1, \ldots, {\sfM}_k)}$.
    Accept if all the qubits in ${\sfV}$ are in state $\ket{0}$,
    and reject otherwise.
  \end{step}
\end{step}
\end{algorithm*}
\caption{Verifier's protocol to reduce the number of turns to two (direct construction).} \label{Figure:
verifier's protocol in one-round system (direct construction)}
\end{figure}

{\em Completeness:} This follows  immediately from the completeness of the proof system in
Lemma~\ref{Theorem: making 3-turn systems public-coin}: in Lemma~\ref{Theorem: making 3-turn systems
public-coin} the first prover sends both ${\sfV}$ (before receiving the bit $b$) and ${\sfM_1}$ (after); here
we can imagine that before the protocol starts the first prover gives register ${\sfV}$ to the extra
$(k+1)$-st prover, who sends it to ${\sfV}$.

{\em Soundness:} This also follows from  the soundness of the proof system in Lemma~\ref{Theorem: making
3-turn systems public-coin}: by combining the actions of the first prover and the extra $(k+1)$-st prover
(and thus making the provers only stronger), we can construct a set of provers that would succeed in the
proof system of Lemma~\ref{Theorem: making 3-turn systems public-coin} with the same probability as they
succeed here.
\end{proof}

\end{document}